\documentclass[aps,pra,epsfigure,twocolumn,superscriptaddress]{revtex4-1}
\usepackage[colorlinks=true,linkcolor=blue,urlcolor=blue,citecolor=blue,pdfusetitle]{hyperref}
\usepackage[utf8]{inputenc}
\usepackage[english]{babel}
\usepackage{amsmath}
\usepackage[caption = false]{subfig}
\usepackage{graphicx,epstopdf}
\usepackage{blindtext}
\usepackage[table,xcdraw]{xcolor}
\usepackage{lipsum}
\usepackage{amsfonts}
\usepackage{bbm}
\usepackage{amssymb}
\usepackage{enumerate}
\usepackage{color}
\usepackage{latexsym}
\usepackage{physics}
\usepackage{times,txfonts}

\usepackage{xfrac}

\newcommand{\Dcal}{\mathcal{D}}

\newcommand{\Fcal}{\mathcal{F}}

\newcommand{\Scal}{\mathcal{S}}

\newcommand{\1}{\mathbbm{1}}

\newcommand{\SubFig}[2]{\ref{#1}{\color{blue}#2}}

\definecolor{MyBlue}{RGB}{0, 76, 153}
\definecolor{MyPurple}{RGB}{204, 0, 102}
\definecolor{MyRed}{RGB}{255, 102, 102}

\newcommand{\UFSCar}{Departamento de Física, Universidade Federal de São Carlos, \\Rodovia Washington Luís, km 235 - SP-310, 13565-905 São Carlos, SP, Brazil}

\newcommand{\SU}{Department of Physics, Stockholm University, AlbaNova University Center 106 91 Stockholm, Sweden}

\usepackage{orcidlink}

\begin{document}

\title{Role of parasitic interactions and microwave crosstalk in dispersive control \\of two superconducting artificial atoms}

\author{Alan C. Santos~\orcidlink{0000-0002-6989-7958}}
\email{ac\_santos@df.ufscar.br}
\affiliation{\UFSCar}
\affiliation{\SU}


\begin{abstract}
	In this work we study the role of parasitic interactions and microwave crosstalk in a system of two superconducting artificial atoms interacting via a single-mode coplanar waveguide. Through a general description of the effective dynamics of the atoms, beyond the two-level approximation, we show that the atom selectivity (ability to individually address an atom) is only dependent on the resultant phasor associated to the drives used to control the system. We then exploit the benefits of such a drive-dependent selectivity to describe how the coherent population inversion occurs in the atoms simultaneously, with no interference of residual atom-atom interaction. In this scenario the parasitic interaction works as a resource to fast and high fidelity control, as it gives rise to a new regime of frequencies for the atoms able to suppress effective atom-atom coupling (idling point). To end, we show how an entangling $i$SWAP gate is implemented with fidelity higher than $99\%$, even in presence of parasitic interactions. More than that, we argue that the existence of this interaction can be helpful to speed up the gate performance. Our results open prospects to a new outlook on the real role of such ``undesired" effects in a system of superconducting artificial atoms.
\end{abstract}

\maketitle

\section{Introduction}

Even with the remarkable recent advances in superconducting quantum computation~\cite{Song:17PRL,Gong:19,Arute:19,Wu:21,Gong:21}, high fidelity effective control of artificial atoms is one of the most challenging task in such a platform that needs to be overcome~\cite{Wendin:17,Siddiqi:21}. A number of different undesired phenomena make this achievement hard, like decoherence effects~\cite{Burnett:19}, or systematic errors for control and interactions in general~\cite{ZhaoPeng:22,Willsch:17,Krinner:20,Han:20,Dai:22}, for example. For this reason, several strategies to suppress errors in superconducting qubits have been investigated~\cite{Corcoles:15,Tripathi:22,Chen:21,Zhao:22}. In recent years, different mechanisms to circumvent errors associated to microwave crosstalk and parasitic interaction have been the main focus in different architectures~\cite{Mates:19,Mundada:19,Dai:21,Ni:21,Cai:21,Nuerbolati:22}, since these errors directly affect gate fidelity in regimes where no decoherence affects' the system. In short, one can say that parasitic interactions are undesired couplings between two qubits due to their physical proximity (distance), while crosstalk refers to the inability to perfectly address a single qubit without affecting different qubits in the system due to their proximity in frequency. It means that, while decoherence drastically affects the system when time of a given computation is large enough, crosstalk and parasitic effects can occur on the timescale of single-shot quantum gates~\cite{Sheldon:16,Mundada:19}.

\begin{figure}[t!]
	\includegraphics[width=\columnwidth]{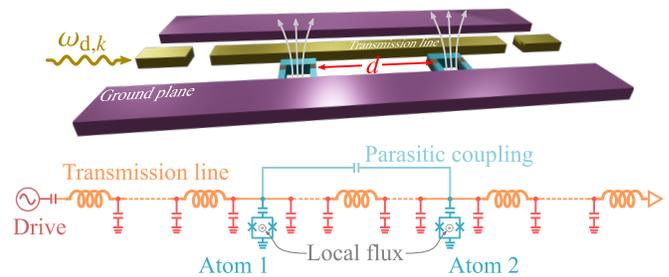}
	\caption{(Top) Two superconducting artificial atoms coupled via a coplanar superconducting waveguide, with a separation $d$ between them. The drive is applied to the transmission line to dispersively control the atoms. We assume frequency-tunable atoms, which can be controlled by a local magnetic flux through the loop of each atom. (Bottom) The circuit representation of the system~\cite{Blais:07}, where we introduce the capacitive coupling expected in cases where the parasitic interaction is not negligible.}\label{Fig:Scheme}
\end{figure}

One of the architectures of superconducting qubits, and main focus of this work, is the system of two transmon qubits coupled to a coplanar waveguide resonator first introduced by Blais \textit{et al}~\cite{Blais:04,Blais:07}, and later experimentally realized by Wallraff \textit{et al}~\cite{Wallraff:04}. As sketched in Fig.~\ref{Fig:Scheme}, two superconducting (transmon) qubits are coupled each other mediated by a (quasi-)one-dimensional coplanar waveguide resonator, in which the distance $d$ separating the qubits is chosen such that no direct interaction between them is detected~\cite{Majer:07,Chow:11}. This kind of system is particularly interesting due to its adaptability (control and readout~\cite{Majer:07,DiCarlo:09,Johnson:10}) and for allowing us to engineer alternative topologies of qubit-qubit coupling~\cite{Leek:10,McKay:19,Cai:21PRR}. For example, similar systems were used to efficiently simulate collective phenomena of Tavis-Cummings model in a strongly interacting three-qubit superconducting system~\cite{Fink:09}, with its scaling up recently reaching the number of $10$ qubits, used to create highly entangled collective super- and sub-radiant states useful to design quantum memories~\cite{Wang:20}. Among others applications collected in a recent review on circuit quantum electrodynamics~\cite{Blais21}.

It is worth mentioning that, even when high controllability is possible for a few qubits system, in cases where many qubits share the same resonator we expect some limitation to single qubit control due to microwave crosstalk~\cite{Nuerbolati:22}, where the qubit frequencies need to be carefully tuned to avoid bringing the system into a chaotic regime~\cite{Berke:22}. More than that, in the case of a finite size circuit, the distance $d$ between the qubits may induce parasitic capacitive couplings and they need to be taken into account in quantum computation~\cite{Xu:20,Chu:22,Luo:22,Bruno:22}. Such interaction is sketched in the circuit of the Fig.~\ref{Fig:Scheme}. In this regard, we will address the task to control the system described in Fig.~\ref{Fig:Scheme} under microwave crosstalk and parasitic interactions, where we work in the regime of operation for the system where none of these effects are negligible. From a general approach of the effective dynamics for two artificial atoms, we show how errors promoted by the microwave crosstalk and parasitic interaction can be mitigated if we have enough information about their nature. To end, as opposite to a negative interpretation of the role of parasitic interactions, we exploit a positive role of such an interaction showing that it is a resource to speed up the physical implementation of the entangling $i$SWAP gate.

\section{The system and effective dynamics}

The Hamiltonian that describes the system considered in this work (atoms, resonator and drive), in the rotating wave approximation, can be written as~\cite{Blais:07,Mitchell:21}
\begin{align}
\hat{H}(t) = \hat{H}_{0} + \hat{H}_{\mathrm{cpg}} + \hat{H}_{\mathrm{d}}(t) , \label{Eq:H_total}
\end{align}
where $\hat{H}_{0}$ is the self-Hamiltonian of the system, where no interaction is taken into account, written as
\begin{align}
	\hat{H}_{0} = \hbar \omega_{\mathrm{r}} \hat{r}^{\dagger}\hat{r} + \hbar \sum\nolimits_{n=1,2} \left(\omega_{n}\hat{a}_{n}^{\dagger}\hat{a}_{n} + \frac{\alpha_{n}}{2}\hat{a}_{n}^{\dagger}\hat{a}_{n}^{\dagger}\hat{a}_{n}\hat{a}_{n}\right) , \label{Eq:H_0}
\end{align}
with $\hat{r}$ ($\hat{r}^{\dagger}$) the annihilation (creation) operator for the resonator of frequency $\omega_{\mathrm{r}}$, $\hat{a}_{n}$ ($\hat{a}_{n}^{\dagger}$) the annihilation (creation) operator for the $n$-th atom with frequency $\omega_{n}$ and anharmonicity parameter $\alpha_{n}$ (considered here positive). The Hamiltonian for the interactions in the system is
\begin{align}
	\hat{H}_{\mathrm{cpg}} = \hbar \sum\nolimits_{k=1}^{2} g_{k} \left(\hat{a}_{k}^{\dagger}\hat{r} + \hat{a}_{k}\hat{r}^{\dagger}\right) + \hbar g \left(\hat{a}_{1}^{\dagger}\hat{a}_{2} + \hat{a}_{1}\hat{a}_{2}^{\dagger}\right) , \label{Eq:H_c}
\end{align}
in which the first term is the interaction between the resonator and each atom, with coupling strength $g_{k}$ for the $k$-th atom. The second one is included here in order to take into account capacitively atom-atom parasitic interactions. It is quite common to neglect such an interaction since its strength is much smaller than the atom-resonator coupling ($|g|\ll |g_{k}|$). For the sake of comparison, some experimental setups are built in such way that parasitic interactions satisfy $|g| \sim |g_{k}|/100$~\cite{Wang:20}. The last term in Eq.~\eqref{Eq:H_total} is the drive applied to the resonator to control the atoms. We assume here the generic case where a number $N$ of drives can be simultaneously applied to the resonator, with different frequencies $\omega_{\mathrm{d},k}$, time-dependent amplitude $\varepsilon_{\mathrm{d},k}(t)$ and phases $\phi_{\mathrm{d},k}$, such that 
\begin{align}
	\hat{H}_{\mathrm{d}}(t) = \sum\nolimits_{k=1}^{N}\hbar\varepsilon_{\mathrm{d},k}(t)\left( \hat{r}^{\dagger}e^{-i\omega_{\mathrm{d},k} t + i\phi_{\mathrm{d},k}} + \hat{r}e^{i\omega_{\mathrm{d},k} t - i \phi_{\mathrm{d},k}}\right) .
\end{align}

It is possible to verify that our description is not complete, as we do not care about dissipation in the system. However, as we aim to determine how to implement fast and high controllable gates, this model is sufficiently useful whenever the time to perform operations is much shorter than the decoherence time scale. Having said that, the time-dependent Schrödinger equation
$i\hbar\ket*{\dot{\psi}(t)}  = \hat{H}(t) \ket*{\psi(t)} $ rules the system evolution. Throughout this work any exact (numerical) simulation of the system is done by the complete Hamiltonian given in Eq.~\eqref{Eq:H_total}, where atoms and resonator have a Hilbert space with dimension $D_\mathrm{atoms}=D_\mathrm{res}=5$ to make sure we are taking into account any effect of the atoms anharmonicity in the effective description of the system.

By following the standard method to describe the system dynamics through a more appropriated approach, considered by Blais \textit{et al}~\cite{Blais:07} for the particular case of a single atom and a single drive resonator, we use a transformation in the resonator operator as given by the time-dependent displacement operator $\hat{\Dcal}(t) = \exp\!~(\xi(t) \hat{r}^{\dagger} - \xi^{\ast}(t) \hat{r})$. This transformation is useful to effectively describe the qubit control through a coherent pumping in the waveguide. In this way, the dynamics of the system is governed by the modified Schrödinger equation $i\hbar\ket*{\dot{\phi}(t)}  = \hat{H}_{\Dcal}(t) \ket*{\phi(t)}$, where we define $\ket*{\phi(t)} = \hat{\Dcal}\ket*{\psi(t)}$ and the new Hamiltonian reads
\begin{align}
	\hat{H}_{\Dcal}(t) = \hat{\Dcal}\hat{H}(t)\hat{\Dcal}^{\dagger} + i\hbar\dot{\hat{\Dcal}} \hat{\Dcal}^{\dagger} .
\end{align}

The additional last term of the modified Hamiltonian is a ``fictitious potential" for non-inertial frames in quantum mechanics introduced by W. Klink~\cite{Klink:97}. Then, applying this result to the Hamiltonian $\hat{H}(t)$ and choosing the free function $\xi(t)$ such that (see Appendix~\ref{ApSec:SemiClass} for further details)
\begin{align}
	i \dot{\xi}(t) =  \omega_{\mathrm{r}}\xi(t) - \sum\nolimits_{k=1}^{N}\varepsilon_{\mathrm{d},k}(t)e^{-i\omega_{\mathrm{d},k} t + i\phi_{\mathrm{d},k}} , \label{Eq:alpha}
\end{align}
one finds
\begin{align}
	\hat{H}_{\Dcal}(t) &= \hat{H}_{0} + \hat{H}_{\mathrm{cpg}} - \hbar \sum\nolimits_{k=1}^{2} g_{k} \left(\hat{a}_{k}^{\dagger} \xi(t) + \hat{a}_{k} \xi^{\ast}(t)\right) ,
\end{align}
in which the last term is the ``semi-classical drive" counterpart $\hat{H}_{\mathrm{dr}}^{\mathrm{cl}}(t)$ of the drive Hamiltonian $\hat{H}_{\mathrm{d}}(t)$. In this way, we can simplify the model because the influence of the drive is directly associated to the atoms operators. Moreover, by solving the Eq.~\eqref{Eq:alpha} we write the final Hamiltonian for the ``semi-classical drive" as
\begin{align}
	\hat{H}_{\mathrm{dr}}^{\mathrm{cl}}(t) &= \hbar \sum_{k=1}^{2}\sum_{n=1}^{N} \Omega_{k,n}(t) \left(\hat{a}_{k}^{\dagger} e^{-i\omega_{\mathrm{d},n}t + i\phi_{\mathrm{d},n}}  + \hat{a}_{k} e^{i\omega_{\mathrm{d},n}t - i\phi_{\mathrm{d},n}} \right) . \label{Eq:HdrCl}
\end{align}
where $\Omega_{k,n}(t) = - g_{k}\varepsilon_{\mathrm{d},n}(t) / \Delta_{R,n}$ is Rabi frequency associated to the drive control over the qubits, with $\Delta_{R,n} = \omega_{\mathrm{r}}-\omega_{\mathrm{d},n}$ the detuning between the resonator frequency and the $n$-th drive field. In some sense, the above equation generalizes the result discussed in the pioneering work on the semi-classical drive derived in Ref.~\cite{Blais:07}. As detailed in Appendix~\ref{ApSec:Drive}, to find the general solution to Eq.~\eqref{Eq:alpha} we do not assume a time-independent drive amplitude $\varepsilon_{\mathrm{d},n}(t)$, as the function $\varepsilon_{\mathrm{d},n}(t)$ is taken as an analytic function within its domain we only need to use the Riemann--Lebesgue lemma~\cite{Serov:17}. As we shall see, the above equation allows us to describe the microwave crosstalk (in this system) between two superconducting qubit driven simultaneously by the resonator.

\subsection{Dispersive control and interaction of the atoms}

In general the frequency of the resonator $\omega_{\mathrm{r}}$ is far from resonance with the atoms $\omega_{n}$, situation in which the \textit{dispersive control} of superconducting qubits takes place. Whenever the detuning $|\omega_{\mathrm{r}}-\omega_{n}|$ is much bigger than the atom-resonator coupling strength $|g_{k}|$, the dynamics of the system is well dictated by an effective Hamiltonian approach. For simplicity, at this moment, the two-level approximation is done in the atoms by replacing the bosonic operators $\hat{a}_{k}$ by the fermionic ones $\hat{\sigma}_{k}^{-}$. Nonetheless, by doing that before the effective Hamiltonian analysis can lead to incorrect predictions of the real evolution for more than one atom in the system~\cite{Bruno:22}. In this regard, we will follow a different methodology to get the Hamiltonian that describes the effective control for the atoms, as well as tunable atom-atom interaction, beyond the two-level approximation. To this end, given only the Hamiltonian without drive $\hat{H}_{\Dcal}^{\mathrm{ndr}} = \hat{H}_{0} + \hat{H}_{\mathrm{cpg}}$, we can ``remove" the coupling of the atoms with the resonator by applying the transformation $\hat{R} = \exp\!~[\hat{S}(\eta)]$, with
\begin{align}
	\hat{S}(\eta) = \eta_{1} (\hat{a}_{1}^{\dagger}\hat{r} - \hat{a}_{1}\hat{r}^{\dagger}) + \eta_{2} (\hat{a}_{2}^{\dagger}\hat{r} - \hat{a}_{2}\hat{r}^{\dagger}) , \label{Eq:STranf}
\end{align}
where $\eta_{k}$'s need to be chosen in order to eliminate the atom-resonator interactions in second order of the Baker-Campbell-Hausdorff expansion
\begin{align}
	\hat{R} \hat{H}^{\mathrm{ndr}}_{\Dcal}\hat{R}^{\dagger} \approx \hat{H}_{\Dcal}^{\mathrm{ndr}} + [\hat{S},\hat{H}_{\Dcal}^{\mathrm{ndr}}] + \frac{1}{2!}\left[\hat{S}, [\hat{S},\hat{H}_{\Dcal}^{\mathrm{ndr}}] \right] . \label{Eq:BCHexp}
\end{align}

By doing this, it is possible to show that the free parameters $\eta_{k}$ need to be (see Appendix~\ref{ApSec:Effective} for further details)
\begin{align}
	\eta_{1} = \frac{g g_{2}+\Delta_{2} g_{1}}{g^2-\Delta_{1} \Delta_{2}} , \quad 
	\eta_{2} = \frac{g g_{1}+\Delta_{1} g_{2}}{g^2-\Delta_{1} \Delta_{2}} ,
\end{align}
where $\Delta_{n} = \omega_{\mathrm{r}} - \omega_{n}$ is the $n$-th atom-resonator detuning. It is worth mentioning here that our choice of $\eta_{k}$ recovers its usual form~\cite{Blais:07,Bravyi:11,Yan:18,Li:20,Feng:20} in limit where the parasitic coupling is completely negligible ($g=0$). Through this simple transformation, the effects of the frequency shift in the resonator (due to the atoms), in the atoms as well (due to the resonator), and the effective atom-atom interaction is observed. In fact, the effective Hamiltonian $\hat{H}_{\mathrm{eff}} =\hat{R} \hat{H}_{\Dcal}^{\mathrm{ndr}}\hat{R}^{\dagger}$, yields
\begin{align}
	\hat{H}_{\mathrm{eff}} &= \hbar \tilde{\omega}_{\mathrm{r}} \hat{r}^{\dagger}\hat{r} + \hbar \sum\nolimits_{n=1,2} \left(\tilde{\omega}_{n}\hat{a}_{n}^{\dagger}\hat{a}_{n} + \frac{\alpha_{n}}{2}\hat{a}_{n}^{\dagger}\hat{a}_{n}^{\dagger}\hat{a}_{n}\hat{a}_{n}\right) \nonumber \\&+
	\hbar g_{\mathrm{eff}} \left(\hat{a}_{1}^{\dagger}\hat{a}_{2} + \hat{a}_{1}\hat{a}_{2}^{\dagger}\right)  , \label{Eq:H_eff}
\end{align}
with the shifted frequencies written as $\tilde{\omega}_{\mathrm{r}} = \omega_{\mathrm{r}} - g_{1}\eta_{1} - g_{2}\eta_{2}$ and $\tilde{\omega}_{n} = \omega_{n} + g_{n}\eta_{n}$, and the effective atom-atom interaction strength mediated by the resonator
\begin{eqnarray}
	g_{\mathrm{eff}} = g + \frac{g \left(g_{1}^2+g_{2}^2\right)+g_{1}g_{2} (\Delta_{1} +\Delta_{2} )}{2(g^2-\Delta_{1} \Delta_{2})}  . \label{Eq:g_eff}
\end{eqnarray}

We are omitting an additional term $\hat{H}_{\mathrm{eff},\alpha}$ related to the effective contribution of the anharmonic part of the atoms (see Appendix~\ref{ApSec:Anharmonicity} for further information). In fact such a term needs to be taken into account in general, but in some particular cases it can be neglected. More precisely, in the regime of \textit{low} excitation in the resonator and when one of the atoms is not doubly excited, the probability of such a term affecting the dynamics is negligible. Consequently, violation of these approximations leads to an effective impact of the $\hat{H}_{\mathrm{eff},\alpha}$ in the system evolution. Its main contribution is associated with a coherent population leakage from the atoms to the resonator, and vice versa, which undermines the fidelity of quantum gates for superconducting \textit{qutrits} and \textit{qudits}, for example.

\subsection{Suppressing parasitic and residual interaction}

Now we discuss how the precise control of a single atom is affected by microwave crosstalk and/or residual two-atoms interactions, and this task can be worse when one uses a single resonator to control more than one atom. When no parasitic interaction affect the system ($g=0$) the local control of the atoms can be efficiently done by driving the system at the \textit{idling point}, that is, a regime of frequencies in which the effective atom-atom interaction is precisely zero. From Eq.~\eqref{Eq:g_eff}, one can see that such a point is achieved for $\Delta_{1} = - \Delta_{2}$, only valid in absence of parasitic interactions. In this context the idling point requests the atom 1 to be quite far from resonance to the atom 2, once the large detuning hypothesis $|\Delta_{k}| \gg |g_{k}|$ needs to be satisfied in the dispersive control. For some quantum operations it is mandatory (or at least desired) to quickly bring the atoms into resonance each other to implement entangling gates avoiding frequency crossing with the resonator, then it is convenient to find a regime in which $\Delta_{1} \neq - \Delta_{2}$. To this end, we use values of frequencies to satisfy $\Delta_{1}\Delta_{2} \gg g_{1}g_{2} (\Delta_{1} +\Delta_{2} )$, where a small (residual) effective interaction $g_{\mathrm{res}} = - g_{1}g_{2} (\Delta_{1} +\Delta_{2} )/2\Delta_{1}\Delta_{2}$ will affect the system (even in the free-parasitic interaction case, $g=0$). There is experimental support to assume that for fast gate implementations this interaction is totally negligible~\cite{Majer:07}. But, for degenerated atoms ($\omega_{1} = \omega_{2}$), residual coupling is always relevant whenever the time required to execute a given computation ($\Delta t_{\mathrm{c}}$) is of order of $\Delta t_{\mathrm{leak}} = \pi/2|g_{\mathrm{res}}|$, with $\Delta t_{\mathrm{leak}}$ the time interval for atom-atom population leakage.

\begin{figure*}
	\centering
	\includegraphics[width=0.95\linewidth]{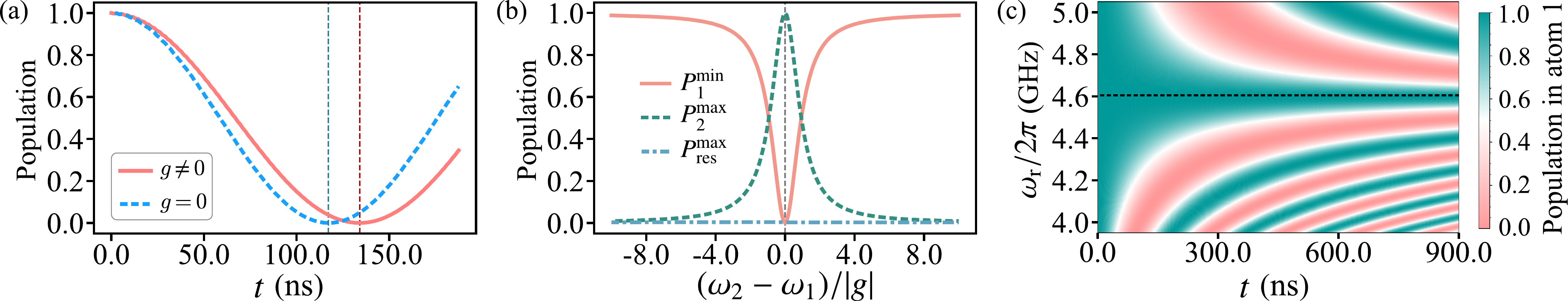}
	\caption{(a) Population dynamics for the atom 1 for $g=0$ and $g = g_{k}/20$ as a function of the time. The vertical dashed line denotes the expected time in which one gets full population leakage  for each model, analytically determine as $\Delta t = \pi / 2g_{\mathrm{eff}}$ with $g_{\mathrm{eff}}$ given in Eq.~\eqref{Eq:g_eff}. The interval between the two vertical lines is $8.7$~ns. 
	(b) Quantities $P_{1}^{\mathrm{min}}$, $P_{2}^{\mathrm{max}}$ and $P_{\mathrm{res}}^{\mathrm{max}}$ as function of $\kappa = (\omega_{2}-\omega_{1})/|g|$. 
	(c) Chevron patter for an excitation initially stored in the atom 1. The horizontal line represents the frequency of the resonator that turns off the atom-atom effective interaction, extracted from Eq.~\eqref{Eq:Omega_rIdle}. 
	For all the graphs we choose $g_{k}/2\pi = 80$~MHz, $\alpha_{k}/2\pi = 300$~MHz, $g=g_1/20$, $\omega_{1}/2\pi = 3$~GHz, and $\omega_{\mathrm{r}}/2\pi = 6$~GHz. For (a) we set $\omega_{2} = \omega_{1}$, and for (b) we choose $\omega_{2}$ as shown on the horizontal axis.}\label{Fig:PopLeak}
\end{figure*}

It is intuitive to guess that population leakage becomes faster for the case of parasitic interaction, since it is an additional coupling. However, as shown in Fig.~\SubFig{Fig:PopLeak}{a} for a particular choice of the frequencies for atoms and resonator, we see a different behavior of the system with the population transfer getting slower for $g\neq0$ than the case with $g=0$. Also, it is seen in Fig.~\SubFig{Fig:PopLeak}{b} that the population leakage can be attenuated by putting the atoms slightly far from resonance, where we assume the same strength of the parasitic coupling as in Fig.~\SubFig{Fig:PopLeak}{a}. By encoding the detuning $\omega_{2} - \omega_{1} = \kappa |g|$, we start the evolution with an excitation in the atom $1$ and evaluate the minimum population of this atom $P_{1}^{\mathrm{min}}$ achieved in the time interval $t\in [0, 10 \times (\pi/2|g_{\mathrm{eff}}|)]$ for different values of $\kappa$. The maximum population in atom 2 ($P_{2}^{\mathrm{max}}$) and in the resonator ($P_{\mathrm{res}}^{\mathrm{max}}$) for the same interval help us to conclude that we have a coherent population transfer to the atom $2$, with no escape to the resonator. More precisely, it is possible to realize that $|\omega_{2} - \omega_{1}| \sim 10|g_{\mathrm{leak}}|$ is enough to get high fidelity for excitation trapping in the atom 1.

Furthermore, the Eq.~\eqref{Eq:g_eff} suggests an alternative to the composition of frequencies for atoms and resonator, in which a protocol robust to population leakage is achievable far from the condition $\Delta_{1} = - \Delta_{2}$, valid in cases where parasitic effects are not negligible. In fact, even for resonant atoms we can avoid the effect seen in Fig.~\SubFig{Fig:PopLeak}{b} if the resonator frequency is suitably chosen to be ($\omega_{1}=\omega_{2}=\omega$)
\begin{eqnarray}
	\omega_{\mathrm{r}}^{\mathrm{idle}} = \omega + \frac{\sqrt{\left(2 g^2+g_{1}^2\right) \left(2 g^2+g_{2}^2\right)} +g_{1} g_{2}}{2 g} . \label{Eq:Omega_rIdle}
\end{eqnarray}
Under this choice, one observes that the parasitic and residual interaction do not provide any contribution to the system dynamics. To check the approximation under which we found such an equation, in Fig.~\SubFig{Fig:PopLeak}{c} we present the Chevron pattern of an excitation initially stored in atom 1. The graph exhibits the time evolution of the population (horizontal axis) as function of the resonator frequency (vertical axis). We highlight the value of the resonator frequency $\omega_{\mathrm{r}}$ able to make Rabi oscillations ``frozen", found through Eq.~\eqref{Eq:Omega_rIdle}. This value of frequency $\omega_{\mathrm{r}}$ will be relevant to single qubit gate implementations.

\section{Single and two qubit gates}

Now we will use the previous discussion to efficiently implement single and two qubit gates in this system. So far the system has been considered as artificial atoms, where the anharmonicity is taken into account. But, at this stage, we will approximate the effective Hamiltonian of the atoms to two-level systems approach by writing
\begin{align}
	\hat{H}_{\mathrm{qubit}} &= \hbar \left(\tilde{\omega}_{1}\hat{\sigma}_{1}^{+}\hat{\sigma}_{1}^{-} + \tilde{\omega}_{2}\hat{\sigma}_{2}^{+}\hat{\sigma}_{2}^{-}\right) +
	\hbar g_{\mathrm{eff}} \left(\hat{\sigma}_{1}^{+}\hat{\sigma}_{2}^{-} + \hat{\sigma}_{1}^{-}\hat{\sigma}_{2}^{+}\right) \nonumber \\
	&+\hbar \sum\nolimits_{k=1}^{2}\Omega_{k}(t) \left(\hat{\sigma}_{k}^{+} e^{-i\omega_{k}t + i\phi_{k}}  + \hat{\sigma}_{k}^{-} e^{i\omega_{k}t - i\phi_{k}} \right) , \label{Eq:H_qubit}
\end{align}
where we omit the resonator operators, as we will work in the highly dispersive regime. In the last term of the above equation we used the phasor addition, which suggests that each qubit will be effectively driven by effective pumping found through the method of phasors as
\begin{align}
\Omega_{k}(t)e^{-i\omega_{k}t + i\phi_{k}} = \sum\nolimits_{n=1}^{N}\Omega_{k,n}(t)e^{-i\omega_{\mathrm{d},n}t + i\phi_{\mathrm{d},n}} ,
\end{align}
with $\Omega_{k}(t)$, $\omega_{k}$ and $\phi_{k}$ the amplitude, frequency and phase of the resultant phasor. This equation clarifies a kind of \textit{crosstalk} effect between the qubits, since the resulting frequency $\omega_{k}$ and phase $\phi_{k}$ depend on the same drive frequencies $\omega_{\mathrm{d},n}$ and phases $\phi_{\mathrm{d},n}$ for the oscillating term of the drives through the resonator. For instance, if a single drive signal ($N=1$) is used to control the qubits, with frequency $\omega_{\mathrm{d}}$ and phase $\phi_{\mathrm{d}}$, one immediately concludes that both atoms will see an oscillating field with frequency $\omega_{k} = \omega_{\mathrm{d}}$ and phase $\phi_{k} = \phi_{\mathrm{d}}$, even if they are driven by fields with different Rabi frequencies $\Omega_{k}(t)$. To better understand how this kind of crosstalk affects the independent control of the qubits, here we will focus on the case of a single drive, such that we can write the Hamiltonian in an oscillating frame, at frequency $\omega_{\mathrm{d}}$, as (see Appendix~\ref{ApSec:LandauZener} for further details)
\begin{align}
	\hat{H}^{\prime} &= \hat{H}_{\mathrm{LZ}}^{(1)} + \hat{H}_{\mathrm{LZ}}^{(2)} +
	\hbar g_{\mathrm{eff}} \left(\hat{\sigma}_{1}^{+}\hat{\sigma}_{2}^{-} + \hat{\sigma}_{1}^{-}\hat{\sigma}_{2}^{+}\right) , \label{Eq:H_prime}
\end{align}
where we define the Landau-Zener like Hamiltonian for each Hamiltonian as ($\tilde{\Delta}_{1}=\tilde{\omega}_{1}-\omega_{\mathrm{d}}$)
\begin{align}
	\hat{H}_{\mathrm{LZ}}^{(k)} = \hbar \tilde{\Delta}_{k}\hat{\sigma}_{k}^{+}\hat{\sigma}_{k}^{-} + \hbar \Omega_{k}(t) \left(\hat{\sigma}_{k}^{+} e^{i\phi_{\mathrm{d}}}  + \hat{\sigma}_{k}^{-} e^{- i\phi_{\mathrm{d}}} \right)
	.
\end{align}

The trivial case where qubits have a huge difference in frequencies, at most one Landau-Zener Hamiltonian will dominate the dynamics and we can decide which one to ``activate" it using the drive frequency $\omega_{\mathrm{d}}$. In fact, when the detuning $|\omega_{\mathrm{d}} - \omega_{k}|$ is much bigger than the Rabi frequency $\Omega_{k}(t)$, then $\hat{H}_{\mathrm{LZ}}^{(k)}\approx \hbar \tilde{\Delta}_{k}\hat{\sigma}_{k}^{+}\hat{\sigma}_{k}^{-}$ and only a local phase will affect the dynamics of the $k$-th qubit, with no population inversion in the computational basis $\ket{0}$ and $\ket{1}$. Of course, assuming that the parasitic and effective coupling strength does not affect the system ($g_{\mathrm{eff}}=0$). However, even if there is parasitic interaction in the system we can choose the resonator frequency $\omega_{\mathrm{r}}$ to bring the system to an idle point and the local control can be done with good fidelity, as wee shall see now.

\subsection{Single qubit gate selectivity}

\begin{figure}[t!]
	\includegraphics[width=\columnwidth]{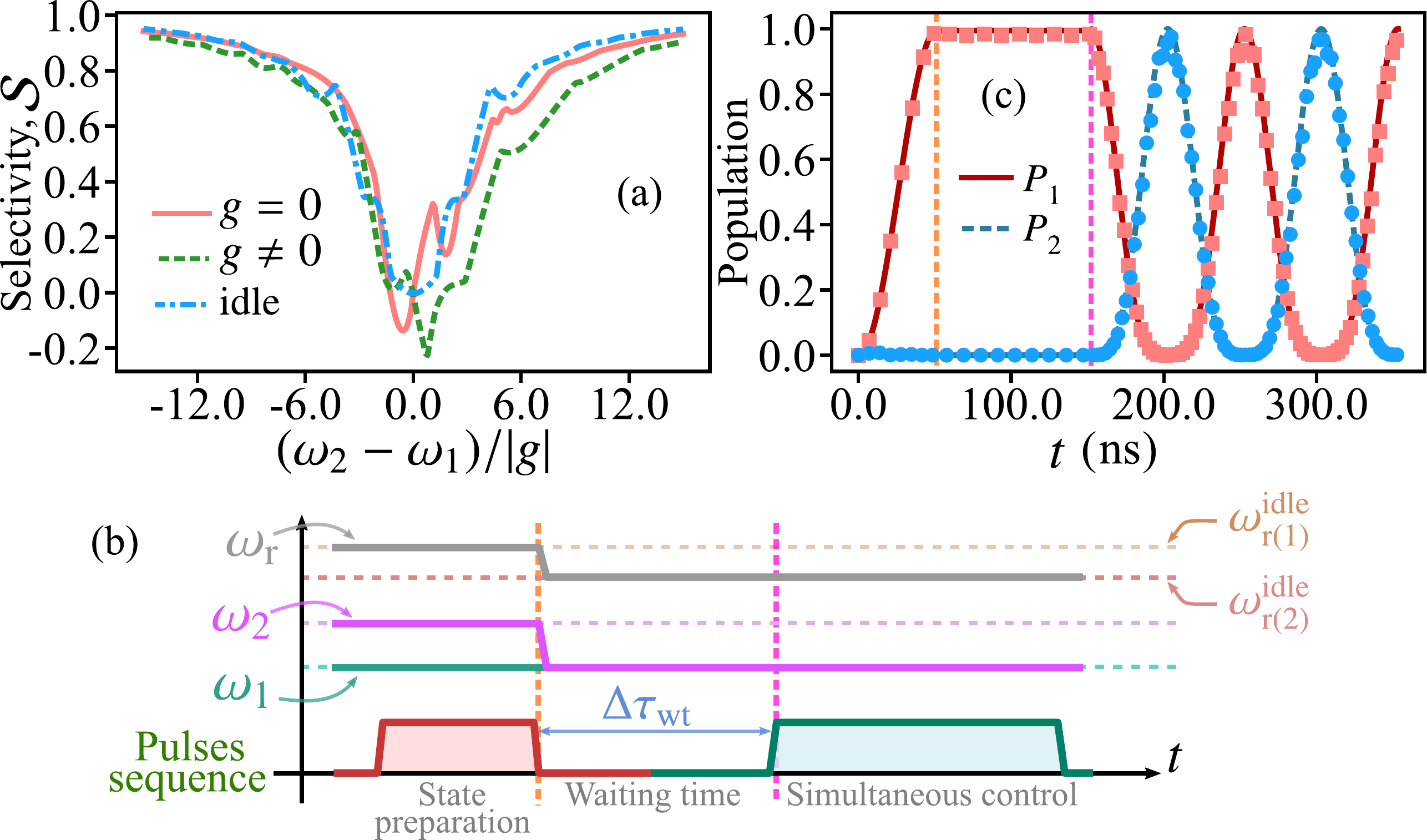}
	\caption{(a) Selectivity as function of the parameter $\kappa$ for different configurations of the system. First, we investigate the role of parasitic interactions by comparing the cases $g = 0$ and $g \neq 0$, where the resonator is far from the idle frequency ($\omega_{\mathrm{r}}/2\pi=5.190$~GHz). Also, we consider the same quantity for the effectively non-interacting case by imposing $\omega_{\mathrm{r}} = \omega_{\mathrm{r}}^{\mathrm{idle}}$.
(b) Circuit used to illustrate the simultaneous control of the qubits. Starting from all qubits in the state $\ket{0}$, the state preparation of the qubit 1 in the state $\ket{1}$ is achieved with the frequency of the qubits detuned as $(\omega_{1} - \omega_{2})/2\pi = 100$~MHz and the resonator frequency is chosen at idling point for this configuration, namely $\omega^{\mathrm{idle}}_{\mathrm{r}(1)}/2\pi\approx8.27$~GHz. It guarantees high fidelity to the state preparation. After that, we switch off the pumping field, frequency of the qubit 2 is then changed to resonance with qubit 1 and we regulate the resonator frequency to a new value, $\omega^{\mathrm{idle}}_{\mathrm{r}(2)}/2\pi\approx8.22$~GHz, keeping the system at the idle configuration. During the time interval (waiting time) no population dynamics is observed. To end, we apply the control field to coherently drive the atoms in their respective and independent Rabi oscillations.
(c) The simulation (dots) and analytical (curves) result for the circuit shown in (b). The parameters have been chosen inspired by Ref.~\cite{Majer:07}, where $\omega_{1}/2\pi=
6.617$~GHz for all cases and the amplitude, and the other parameters as given in Fig.~\ref{Fig:PopLeak}. The drive amplitude is taken as $\varepsilon_{\mathrm{d},k}(t)/2\pi = 100$~MHz.}\label{Fig:SingleQubitControl}
\end{figure}

One defines the \textit{selectivity} as a measurement of individual controllability, that is, it quantifies how well we can control one of the atoms, by pumping a transmission line, without affecting the other one. In general it is expected that the selectivity depends on the gate to be implemented, so here we will focus on the perfect control $X$ Pauli gates ($\pi$ pulse). In this case, only the population of the qubit 1 (addressed qubit) starts from zero to one while the population of the atom 2 (spectator atom) should remain zero all the time. Mathematically we can measure this selectivity from the equation
\begin{eqnarray}
\Scal = \frac{P_{\mathrm{add}}^{\mathrm{max}} - P_{\mathrm{spe}}^{\mathrm{max}}}{P_{\mathrm{add}}^{\mathrm{max}} + P_{\mathrm{spe}}^{\mathrm{max}}} ,
\end{eqnarray}
with $P_{\mathrm{add}}^{\mathrm{max}}$ and $P_{\mathrm{spe}}^{\mathrm{max}}$ denoting the maximal population induced by the drive in the addressed and spectator atoms, respectively. 
This quantity ranges from $-1$ to $1$, which corresponds to the \textit{inversely} selective ($P_{\mathrm{add}}^{\mathrm{max}}=0$ and $P_{\mathrm{spe}}^{\mathrm{max}}=1$) to the \textit{completely} selective regime ($P_{\mathrm{add}}^{\mathrm{max}}=1$ and $P_{\mathrm{spe}}^{\mathrm{max}}=0$). We also highlight the case $\Scal =0$, which corresponds to the scenario of equal controllability of the qubits. It means that, at this point, we can simultaneously execute (identical) operations on the atoms. The Fig.~\SubFig{Fig:SingleQubitControl}{a} reports the selectivity $\Scal$ for the qubit $1$ as addressed qubit for three different cases as function of the qubits detuning. The first two cases take into account the effect of parasitic interactions in the system (solid red curve and dashed green line, respectively). As expected, in all situations we see low selectivity at (or close to) resonance between the atoms, but it becomes even worse when parasitic interactions are present in the system. Also we exploit the counter-intuitive benefit of having parasitic interaction in the system, by showing the behavior of the selectivity evolving the system at the idling point (dot-dashed blue line). When the qubits are tuned at resonance, the idling point shows that the selectivity is close to zero, which means the simultaneous control of both atoms would be possible. To demonstrate such a property, we implement the sequence of pulse and frequencies as sketched in Fig.~\SubFig{Fig:SingleQubitControl}{b}, as detailed below.

A single driving field is applied to the two-qubit system through the resonator, with drive strength $\varepsilon_{\mathrm{d},k}(t) = \varepsilon_{0}$ (consequently $\Omega_{k}=\Omega$) and phase $\phi_{k} = 0$. The simultaneous coherent control of the qubits is properly investigated by preparing the qubit $1$ in state $\ket{1}$, through a $\pi$-pulse, while the qubit 2 starts in the ground state. To this end, the qubits and resonator are far from resonance, and the drive is assumed to be at resonance with the qubit 1 ($\omega_{\mathrm{d}} = \tilde{\omega}_{1}$). During the step of state preparation the excited state population in the qubit 1 evolves as $P_{\mathrm{exc}}^{(1)}(t) = \sin^2(\Omega t)$, while the qubit 2 remains in its ground state, as provided by the analytical solution of the Eq.~\eqref{Eq:H_prime}. Here we approximate the solution to the case with effective interaction $g_{\mathrm{eff}}=0$, since such an interaction is much smaller than the Rabi frequency $|\Omega|$. After that, to highlight the idling point performance (already shown in Fig.~\SubFig{Fig:PopLeak}{c}), we align the qubits frequencies, $\omega_{1}=\omega_{2}$, and set the resonator at the idle frequency. During this ``waiting" step, where the pumping is turned off $(\Omega \propto \varepsilon_{0}=0)$, we verify that no population transfer is done, which we expect whenever the perfect suppression of effective interaction in the system is used, including parasitic ones. In the last step of this theoretical simulation we apply a driving field to induce identical Rabi oscillations in both qubits, since the qubits are in resonance each other. For this step the instantaneous population in the excited state for the qubits 1 and 2 are given, respectively, by $P_{\mathrm{exc}}^{(1)}(t) = \cos^4(\Omega t)$ and $P_{\mathrm{exc}}^{(2)}(t) = \sin^4(\Omega t)$ (due to their initial states). In Fig.~\SubFig{Fig:SingleQubitControl}{c} we present the analytical predictions for the populations and the exact numerical simulation from the Hamiltonian in Eq.~\eqref{Eq:H_total}. The agreement between analytical results, which presumes the perfect absence of effective interaction, and the numerical simulation (using QuTiP~\cite{QuTiP-1,QuTiP-2}) illustrates the pertinence of the effective description of the system introduced in this work.

It is worth now to mention the experimental implementation, by Majer \textit{et al}~\cite{Majer:07}, that supports our theoretical discussion. Using a system of two coupled qubits via a cavity bus, in such a work the authors showed the absence of beating in the dynamics for population of the qubits, indicating that the coupling does not affect the system. To this end, the configuration of frequencies $\Delta_{1}\Delta_{2} \gg g_{1}g_{2} (\Delta_{1} +\Delta_{2} )$ is used and the atom-atom detuning is $|\omega_{1}-\omega_{2}|\approx 88$~MHz. From our analytical results for the effective dynamics, we use the same parameters as considered in such a experimental work to estimate that a difference $|\omega_{1}-\omega_{2}| \geq 80$~MHz would be enough to have a similar result. This discussion becomes relevant here because we can use these values to theoretically evaluate the existence of parasitic interactions in such a system. In fact, from the behavior of the quantity $|\omega_{1}-\omega_{2}|$ as function of the coupling $g$ shown in Fig.~\SubFig{Fig:SingleQubitControl}{c}, one realizes that the selectivity reported in Ref.~\cite{Majer:07} should be affected in case of a parasitic coupling strength $g/2\pi > 1$~MHz.

\subsection{Speeding up iSWAP$^\dagger$ gate using parasitic coupling}

\begin{figure*}[t!]
	\centering
	\includegraphics[width=0.9\linewidth]{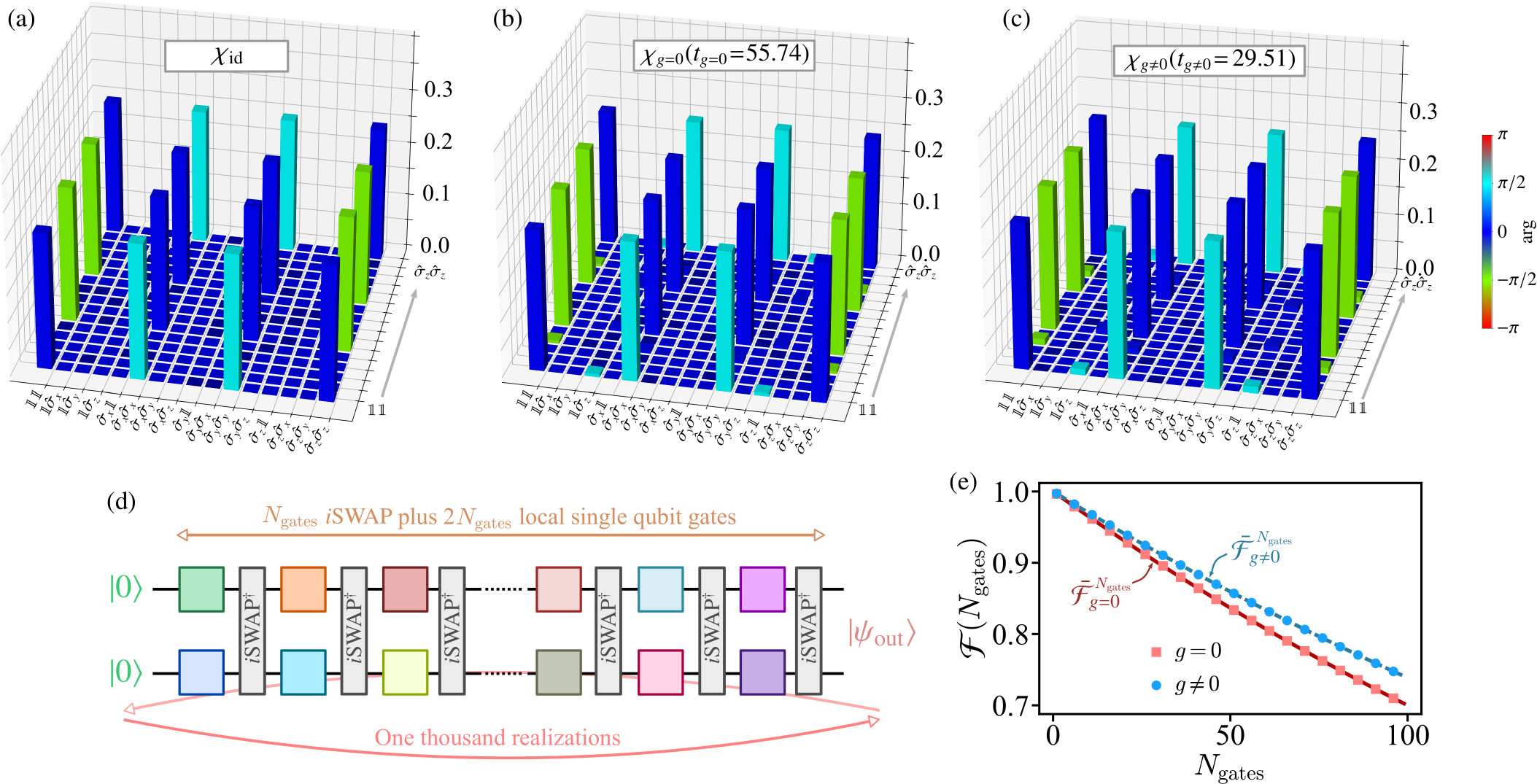}
	\caption{(a) The ideal process matrix $\chi$ for the $i$SWAP$^\dagger$ gate, (b) for $g=0$ obtained for the total evolution time of $t_{g=0}=55.74$~ns, and (c) for $g\neq0$ obtained for the time interval $t_{g\neq0}=29.51$~ns.
		(d) The circuit used in the randomized benchmarking of the $i$SWAP$^\dagger$ gate. (e) The average fidelity obtained in the randomized benchmarking as function of the number of gates considered in the circuit. The frequencies used here are $\omega_{\mathrm{r}}/2\pi=5.190$~GHz and $\omega_{k}/2\pi=6.617$~GHz for the qubits, while the other parameters are the same as in Fig.~\ref{Fig:PopLeak}.}\label{Fig:MatrixProcess}
\end{figure*}

Two qubit gates in superconducting qubits are implemented using the Hamiltonian as given in Eq.~\eqref{Eq:H_qubit}, as due to the hoping term we can control the excitation flux between the qubits. In the particular system considered here we assume the analytical evolution operator (propagator) for the Hamiltonian in Eq.~\eqref{Eq:H_qubit} given by
\begin{eqnarray}
\hat{U}(t) = \exp(i \hat{H}_{\mathrm{pha}} t/\hbar)\exp(-i \hat{H}_{\mathrm{int}} t/\hbar) , \label{Eq:Propagator}
\end{eqnarray}
where $\hat{H}_{\mathrm{int}}=\hbar g_{\mathrm{eff}} \left(\hat{\sigma}_{1}^{+}\hat{\sigma}_{2}^{-} + \hat{\sigma}_{1}^{-}\hat{\sigma}_{2}^{+}\right)$ is the swapping term of the dynamics and $\hat{H}_{\mathrm{pha}}=\hbar \left(\tilde{\omega}_{1}\hat{\sigma}_{1}^{+}\hat{\sigma}_{1}^{-} + \tilde{\omega}_{2}\hat{\sigma}_{2}^{+}\hat{\sigma}_{2}^{-}\right)$ is responsible for generating quantum phases in the system, but it can be suitably corrected by local phase gates applied to the output state. It is possible to show that if we let the system evolves for an interval of time $\tau_{\mathrm{SWAP}} = t = \pi / 2g_{\mathrm{eff}}$, one gets
\begin{align}
\exp(-i \hat{H}_{\mathrm{int}} \tau_{\mathrm{SWAP}}/\hbar) = \left(
\begin{array}{cccc}
	1 & 0 & 0 & 0 \\
	0 & 0 & -i & 0 \\
	0 & -i & 0 & 0 \\
	0 & 0 & 0 & 1 \\
\end{array}
\right),
\end{align}
which corresponds to the Hermitian conjugate of the $i$SWAP gate. Therefore, we can write (up to a global phase) 
\begin{align}
	\hat{U}(\tau_{\mathrm{SWAP}}) = \left[S_{1}(\varphi_{k}) S_{2}(\varphi_{k}) \right] \cdot i\mathrm{SWAP}^{\dagger} , 
\end{align}
where $S_{n}(\varphi_{k})$ are \textit{local} phase shift gates for each atom, with $\varphi_{k} = \pi \tilde{\omega}_{k} /g_{\mathrm{eff}}$, such that we need to implement these local known corrections to the output state. For this operation, the gate time is inversely proportional to the effective interaction strength $g_{\mathrm{eff}}$, which means that such a gate can be positively affected by parasitic interactions in the system. To show this result, we assume the identical coupling strength and atoms at resonance each other to make sure that the hopping term will drive the system through the desired dynamics (up to local error corrections, as already shown). We also take parameters experimentally feasible, with couplings $g_{k}/2\pi = 80$~MHz and frequencies $\omega_{k}/2\pi = 6.617$~MHz. To take advantage of the parasitic coupling we need a negative detuning $\Delta_{k} < 0$, such that parasitic interaction $g$ positively contributes to speed up the dynamics, therefore we choose the resonator frequency $\omega_{\mathrm{r}}/2\pi = 5.19$~GHz. We highlight that such a values of frequency were used in Ref.~\cite{Majer:07}. By using these values we compute the ideal propagator from Eq.~\eqref{Eq:Propagator}, and numerically solve the propagator equation for the Hamiltonian~\eqref{Eq:H_total} (with no pumping) for the cases where the parasitic coupling strength is $g=0$ and $g=g_{k}/20$ (weak parasitic interaction). For each dynamics we also implement the corresponding local phase shift corrections, and the final result obtained through the quantum tomography of the process matrix $\chi$~\cite{Nielsen:Book} is shown in Fig.~\ref{Fig:MatrixProcess}. In the simulation we pick the process matrix $\chi$ evaluated at $\tau_{\mathrm{SWAP}}$, as analytically predicted by Eq.~\eqref{Eq:Propagator}. To obtain the simulated matrix $\chi$ we first find the propagator $\hat{U}_{\mathrm{tot}}(t)$ for the Hamiltonian in Eq.~\eqref{Eq:H_total}, then we project $\hat{U}_{\mathrm{tot}}(t)$ in the two-qubit Hilbert subspace, such that $\chi$ is extracted using the standard matrix tomography~\cite{Nielsen:Book}. The result is shown in Fig.~\SubFig{Fig:MatrixProcess}{a} for the ideal $i$SWAP$^\dagger$, Figs.~\SubFig{Fig:MatrixProcess}{b} and~\SubFig{Fig:MatrixProcess}{c} for the simulation with $g=0$ and $g\neq0$, respectively. For the parameters considered in the simulation, the process matrix shown in Fig.~\SubFig{Fig:MatrixProcess}{b} is obtained by setting the total evolution time as $\Delta t_{g=0} = 55.74$~ns, while the case in Fig.~\SubFig{Fig:MatrixProcess}{c} demands an amount of time $\Delta t_{g\neq0}=29.51$~ns. 

The average fidelity of each dynamics ($g=0$ and $g\neq0$) with respect to the ideal one is addressed through the randomized benchmarking, as sketched in Fig.~\SubFig{Fig:MatrixProcess}{d}. It is desired here to measure the average fidelity $i$SWAP gate, so the circuit is comprised for $N_{\mathrm{gates}}$ $i$SWAP gates interspersed with local single qubit gates. The random property of the circuit is encoded in the generic local qubit gates written as $G_{\mathrm{local}} = G_{1}(\theta_{1},\phi_{1},\lambda_{1}) \otimes G_{2}(\theta_{2},\phi_{2},\lambda_{2})$ with
\begin{align}
G_{k}(\theta_{k},\phi_{k},\lambda_{k}) = \left(
\begin{array}{cc}
	\cos(\theta_{k}/2) & -e^{i\lambda_{k}}\sin(\theta_{k}/2) \\
	e^{i\phi_{k}}\sin(\theta_{k}/2) & e^{i(\lambda_{k}+\phi_{k})}\cos(\theta_{k}/2)  
\end{array}
\right) ,
\end{align}
acting on the $k$-th qubit, and the triplet $\theta_{k}$, $\phi_{k}$ and $\lambda_{k}$ is randomly chosen in the interval $[0,2\pi]$, $[0,\pi]$ and $[0,\pi]$, respectively, for each step of the circuit in Fig.~\SubFig{Fig:MatrixProcess}{d}. 

By measuring fidelity of a single realization from the Bures metric for pure states~\cite{Nielsen:Book}, $\Fcal_{g=0}(\psi,\psi^{\prime}) = | \bra*{\psi}\ket*{\psi^{\prime}} |^2$, we then average the circuit over one thousand of realizations. We define the fidelities $\Fcal_{g=0} = \Fcal(\psi_{\mathrm{ideal}},\psi_{g = 0})$ and $\Fcal_{g \neq 0} = \Fcal(\psi_{\mathrm{ideal}},\psi_{g \neq 0})$ for each dynamics, in which $\ket{\psi_{\mathrm{ideal}}}$ is the ideal expected computation outcome and $\ket*{\psi_{g = 0}}$ ($\ket*{\psi_{g \neq 0}}$) is the simulation output with no parasitic interaction $g = 0$ (with parasitic interaction $g \neq 0$). Fig.~\SubFig{Fig:MatrixProcess}{e} shows the result for a circuit with up to $100$ $i$SWAP gates (plus $100$ pairs of local qubit gates $G_{\mathrm{local}}$). Therefore, by fitting the set of data to determine the average success fidelity $\bar{\Fcal}$ of a $i$SWAP gate after $N_{\mathrm{gates}}$ of the circuit, in Fig.~\SubFig{Fig:MatrixProcess}{d}, according to the fit $\Fcal(N_{\mathrm{gates}}) = \bar{\Fcal}^{N_{\mathrm{gates}}}$, one gets the respective fidelities $\bar{\Fcal}_{g=0} = 99.64(4)$ and $\bar{\Fcal}_{g\neq0} = 99.69(7)$. These values are achieved after many realizations, which means that for a single realization such values will be different. However, the main result after this analysis is that the effects parasitic interaction in such a system can be efficiently suppressed if we know its strength.

\section{Conclusions and prospects}

In this work we discuss strategies able to suppress the negative effects of undesired interactions and atom-atom crosstalk in a system of two interacting superconducting artificial atoms, via a single mode cavity bus, controlled by the same driving channel (transmission line). The high fidelity in the individual coherent control and gate implementations is discussed, with further details of the effective dynamics shown in Appendix section. As main result of this work, we state that the negative role of parasitic interactions and microwave crosstalk needs to be investigated case by case, since we showed how to take advantage of these effects for this system in particular. For instance, the simultaneous control of two qubits is only possible in this system because they feel the same effective resultant phasor propagating through the transmission line. More than that, the additional capacitive interaction between the atoms (avoided in some systems) is used here to enhance the time required to apply $i$SWAP gate in the system, without any negative impact in the average fidelity of the operation.

The strategy considered in this work to highlight the positive impacts of ``undesired" couplings and crosstalk can be, in principle, experimentally investigated. However, because a number of other effects are not taken into account, for instance the non-radiative decay of the atoms and losses in the resonator, a generalization of the effective dynamics examined in this work to open systems needs to be developed. In addition, it is important to highlight that for superconducting artificial atoms with weak anharmonicity, the performance of the $i$SWAP will be affected due to the population leakage to energy levels outside the qubit subspace $\{\ket{0}, \ket{1}\}$ (as detailed in Appendix~\ref{ApSec:WeakAnharmonicity}). Given the recent advance in the control of superconducting platforms to demonstrate the performance of qutrit gates~\cite{Yurtalan:20,Kononenko:21,Luo:22,Goss:22}, the additional coherent atom-resonator coupling induced by the anharmonicity may lead to an important source of gate error to be mitigated. Such a error can also be relevant to different protocols using superconducting qutrits, like holonomic quantum computation in superconducting qubits~\cite{Xue:17,Li:21,Ma:22}, and the charging process of superconducting quantum batteries~\cite{Hu:22a}, for instance.

\begin{acknowledgments}
	A.C.S. acknowledges the financial support of the São Paulo Research Foundation (FAPESP) (Grants No. 2019/22685-1 and 2021/10224-0).
\end{acknowledgments}

\appendix


\section{Proof for the Semi-classical drive equation} \label{ApSec:SemiClass}

Consider the Hamiltonian for system of two artificial atoms and a single driving resonator given by
	\begin{align}
		\hat{H}(t) = \hat{H}_{0} + \hat{H}_{\mathrm{cpg}} + \hat{H}_{\mathrm{d}}(t) , \label{Ap-Eq:H_total}
	\end{align}
	where
	\begin{align}
		\hat{H}_{0} &= \hbar \omega_{\mathrm{r}} \hat{r}^{\dagger}\hat{r} + \hbar \sum\nolimits_{n=1,2} \left(\omega_{n}\hat{a}_{n}^{\dagger}\hat{a}_{n} + \frac{\alpha_{n}}{2}\hat{a}_{n}^{\dagger}\hat{a}_{n}^{\dagger}\hat{a}_{n}\hat{a}_{n}\right) , \label{Ap-Eq:H_0}  \\
		\hat{H}_{\mathrm{cpg}} &= \hbar \sum\nolimits_{k=1}^{2} g_{k} \left(\hat{a}_{k}^{\dagger}\hat{r} + \hat{a}_{k}\hat{r}^{\dagger}\right) + \hbar g \left(\hat{a}_{1}^{\dagger}\hat{a}_{2} + \hat{a}_{1}\hat{a}_{2}^{\dagger}\right) , \label{Ap-Eq:H_c} \\
		\hat{H}_{\mathrm{d}}(t) &= \sum\nolimits_{k=1}^{N}\hbar\varepsilon_{\mathrm{d},k}(t)\left( \hat{r}^{\dagger}e^{-i\omega_{\mathrm{d},k} t + i\phi_{\mathrm{d},k}} + \hat{r}e^{i\omega_{\mathrm{d},k} t - i \phi_{\mathrm{d},k}}\right) .
	\end{align}

Now, as a first approximation, to have a description of a classical pulse applied to the system we change $\hat{H}_{\mathrm{d}}(t)$ for its classical counterpart by introducing the displacement operator rotation. To this end we write the Schrodinger equation driven by the Hamiltonian $\hat{H}(t)$, through the operator $\hat{\Dcal} = \exp (\xi(t) \hat{r}^{\dagger} - \xi^{\ast}(t) \hat{r})$, as
\begin{align}
	i\hbar \hat{\Dcal}\ket*{\dot{\psi}(t)} = \hat{\Dcal}\hat{H}(t)\ket{\psi(t)} .
\end{align}
and we define $\ket{\phi(t)} = \hat{\Dcal}\ket{\Psi(t)}$ to write
\begin{align}
	i\hbar\ket*{\dot{\phi}(t)}  = \hat{H}_{\Dcal}(t) \ket{\phi(t)} .
\end{align}
where
\begin{align}
	\hat{H}_{\Dcal}(t) = \hat{\Dcal}\hat{H}(t)\hat{\Dcal}^{\dagger} + i\hbar\frac{d\hat{\Dcal}}{dt} \hat{\Dcal}^{\dagger} .
\end{align}

By defining the Hamiltonian $\hat{H}_{\mathrm{qb}} = \hat{H}_{\mathrm{int},0} + \hat{H}_{\alpha}$, where the anharmonicity part reads $\hat{H}_{\alpha} = \sum_{n=1,2}(\alpha_{n}/2) \hat{a}_{n}^{\dagger}\hat{a}_{n}^{\dagger}\hat{a}_{n}\hat{a}_{n}$ and
\begin{align}
	\hat{H}_{\mathrm{int},0} = \hbar \sum_{n=1,2} \omega_{n}\hat{a}_{n}^{\dagger}\hat{a}_{n}  +
	\hbar g \left(\hat{a}_{1}^{\dagger}\hat{a}_{2} + \hat{a}_{1}\hat{a}_{2}^{\dagger}\right) ,
\end{align}
we write
\begin{align}
	\hat{H}_{\Dcal}(t) &= \hat{H}_{\mathrm{qb}} + \hbar \omega_{\mathrm{r}} \hat{\Dcal}\hat{r}^{\dagger}\hat{r} \hat{\Dcal}^{\dagger}
	+ \hbar \sum_{k=1}^{2} g_{k} \left(\hat{a}_{k}^{\dagger}\hat{\Dcal}\hat{r}\hat{\Dcal}^{\dagger} + \hat{a}_{k}\hat{\Dcal}\hat{r}^{\dagger}\hat{\Dcal}^{\dagger}\right)
	\nonumber \\
	&+
	\sum_{k=1}^{N}\hbar\varepsilon_{\mathrm{d},k}(t)\left( \hat{\Dcal}\hat{r}^{\dagger} \hat{\Dcal}^{\dagger} e^{-i\omega_{\mathrm{d},k} t + i\phi_{\mathrm{d},k}} + \hat{\Dcal}\hat{r}\hat{\Dcal}^{\dagger} e^{i\omega_{\mathrm{d},k} t - i \phi_{\mathrm{d},k}}\right)
\end{align}

We know that $\hat{\Dcal}\hat{r}\hat{\Dcal}^{\dagger} = \hat{r} - \xi(t)$ and $\left(\hat{\Dcal}\hat{r}\hat{\Dcal}^{\dagger}\right)^\dagger = \hat{\Dcal}\hat{r}^{\dagger}\hat{\Dcal}^{\dagger} = \hat{r}^{\dagger} - \xi^{\ast}(t)$, which allows us to write ($\xi = \xi(t)$ for short)
\begin{align}
	\hat{H}_{\Dcal}(t) &= \hat{H}_{\mathrm{qb}} + \hbar \omega_{\mathrm{r}} \left[\hat{r}^{\dagger}\hat{r}~{\color{red} -~\hat{r}^{\dagger}\xi - \xi^{\ast}\hat{r}} + |\alpha^{\ast}|^{2} \1\right]
	{\color{red}+ i\hbar \left(\dot{\xi} \hat{r}^{\dagger} - \dot{\xi}^{\ast} \hat{r}\right)}
	\nonumber \\
	&
	{\color{red}+
		\sum_{k=1}^{N}\hbar\varepsilon_{\mathrm{d},k}(t)\left( \hat{r}^{\dagger} e^{-i\omega_{\mathrm{d},k} t + i\phi_{\mathrm{d},k}} + \hat{r} e^{i\omega_{\mathrm{d},k} t - i \phi_{\mathrm{d},k}}\right)}
	\nonumber \\
	&-
	\sum_{k=1}^{N}\hbar\varepsilon_{\mathrm{d},k}(t)\left( \xi^{\ast} e^{-i\omega_{\mathrm{d},k} t + i\phi_{\mathrm{d},k}} + \xi e^{i\omega_{\mathrm{d},k} t - i \phi_{\mathrm{d},k}}\right)\1
	\nonumber \\
	&+ \hbar \sum_{k=1}^{2} g_{k} \left(\hat{a}_{k}^{\dagger} (\hat{r} - \xi) + \hat{a}_{k}(\hat{r}^{\dagger} - \xi^{\ast})\right)
	 .
\end{align}

Rearranging the terms in red, one gets
\begin{align}
	\hat{H}_{\Dcal}(t) &= \hat{H}_{\mathrm{qb}} + \hbar \omega_{\mathrm{r}}\hat{r}^{\dagger}\hat{r}  +\hbar \sum_{k=1}^{2} g_{k} \left(\hat{a}_{k}^{\dagger} (\hat{r} - \xi) + \hat{a}_{k}(\hat{r}^{\dagger} - \xi^{\ast})\right) \nonumber \\
	&- \left[
	\sum_{k=1}^{N}\hbar\varepsilon_{\mathrm{d},k}(t)\left( \alpha^{\ast} e^{-i\omega_{\mathrm{d},k} t + i\phi_{\mathrm{d},k}} + \alpha e^{i\omega_{\mathrm{d},k} t - i \phi_{\mathrm{d},k}}\right) - \hbar \omega_{\mathrm{r}}|\alpha^{\ast}|^{2} \right]\1
	\nonumber \\
	&
	{\color{red}+\hat{r}^{\dagger} \left[i\hbar \dot{\xi} - \hbar \omega_{\mathrm{r}}\xi +
		\sum_{k=1}^{N}\hbar\varepsilon_{\mathrm{d},k}(t)e^{-i\omega_{\mathrm{d},k} t + i\phi_{\mathrm{d},k}} \right] }
	\nonumber \\
	&
	{\color{red}	+ \hat{r}\left[
		\sum_{k=1}^{N}\hbar\varepsilon_{\mathrm{d},k}(t) e^{i\omega_{\mathrm{d},k} t - i \phi_{\mathrm{d},k}}
		- i\hbar \dot{\xi}^{\ast}  - \hbar \omega_{\mathrm{r}}\xi^{\ast}\right]} .
\end{align}

Now, one realizes that by choosing $\xi$ such that
\begin{align}
	i \dot{\xi} =  \omega_{\mathrm{r}}\xi - \sum_{k=1}^{N}\varepsilon_{\mathrm{d},k}(t)e^{-i\omega_{\mathrm{d},k} t + i\phi_{\mathrm{d},k}} , \label{Ap-Eq:alpha}
\end{align}
we find
\begin{align}
		\hat{H}_{\Dcal}(t) &= \hat{H}_{0} + \hbar g \left(\hat{a}_{1}^{\dagger}\hat{a}_{2} + \hat{a}_{1}\hat{a}_{2}^{\dagger}\right) + \hat{H}_{\mathrm{int}} + \hat{H}_{\mathrm{dr}} ,
\end{align}
where $\hat{H}_{\mathrm{dr}}$ is the drive and $\hat{H}_{\mathrm{int}}$ is the atoms-resonator interaction part defined, respectively, as
\begin{align}
	\hat{H}_{\mathrm{dr}} &= - \hbar \sum_{k=1}^{2} g_{k} \left(\hat{a}_{k}^{\dagger} \xi + \hat{a}_{k} \xi^{\ast}\right) \label{Ap-Eq:EffDrive}
	\\
	\hat{H}_{\mathrm{int}} &= \hbar \sum_{k=1}^{2} g_{k} \left(\hat{a}_{k}^{\dagger} \hat{r} + \hat{a}_{k}\hat{r}^{\dagger} \right) , 
\end{align}
since the terms proportional to the identity can be included in a zero-point energy shift and do not change the system dynamics. In conclusion, because $\hat{H}_{\mathrm{cpg}} = \hat{H}_{\mathrm{int}} + \hbar g \left(\hat{a}_{1}^{\dagger}\hat{a}_{2} + \hat{a}_{1}\hat{a}_{2}^{\dagger}\right)$ as defined in Eq.~\eqref{Ap-Eq:H_c}, we find
\begin{align}
	\hat{H}_{\Dcal}(t) &= \hat{H}_{0} + \hat{H}_{\mathrm{cpg}} - \hbar \sum\nolimits_{k=1}^{2} g_{k} \left(\hat{a}_{k}^{\dagger} \xi(t) + \hat{a}_{k} \xi^{\ast}(t)\right) .
\end{align}

\section{The Hamiltonian for the driving field in the ``steady-state" regime} \label{ApSec:Drive}

Now, we discuss the final form for the driving field by starting from Eq.~\eqref{Ap-Eq:alpha}. Now, by using
\begin{align}
	e^{-i\omega_{\mathrm{r},k}t}\frac{d}{dt} \left(e^{i\omega_{\mathrm{r},k}t} \xi(t)\right) &= e^{-i\omega_{\mathrm{r},k}t} \left(i\omega_{\mathrm{r},k} e^{i\omega_{\mathrm{r},k}t} \xi(t) + e^{i\omega_{\mathrm{r},k}t} \dot{\xi}(t)\right) \nonumber \\
	&= i\omega_{\mathrm{r},k} \xi(t) + \dot{\xi}(t) ,
\end{align}
one gets
\begin{align}
	e^{-i\omega_{\mathrm{r},k}t}\frac{d}{dt} \left(e^{i\omega_{\mathrm{r},k}t} \xi(t)\right) =  i\varepsilon_{\mathrm{d},k}(t)e^{-i\omega_{\mathrm{d},k} t + i\phi_{\mathrm{d},k}} ,
\end{align}
whose solution reads
\begin{align}
	\xi(t)= \xi(-\infty)e^{i\omega_{\mathrm{r},k}t} + ie^{i\omega_{\mathrm{r},k}t + i\phi_{\mathrm{d},k}}\int_{-\infty}^{t} \varepsilon_{\mathrm{d},k}(t^{\prime})e^{-i(\omega_{\mathrm{d},k}-\omega_{\mathrm{r},k}) t^{\prime}} dt^{\prime}.
\end{align}

Solving the integral
\begin{align}
	\int_{-\infty}^{t} \varepsilon_{\mathrm{d},k}(t^{\prime})&e^{-i(\omega_{\mathrm{d},k}-\omega_{\mathrm{r},k}) t^{\prime}} dt^{\prime} = \left.
	\frac{\varepsilon_{\mathrm{d},k}(t^{\prime})e^{-i(\omega_{\mathrm{d},k}-\omega_{\mathrm{r},k}) t^{\prime}} }{-i(\omega_{\mathrm{d},k}-\omega_{\mathrm{r},k})}\right\vert_{-\infty}^{t} \nonumber \\&-
	\frac{1}{-i(\omega_{\mathrm{d},k}-\omega_{\mathrm{r},k})}\int_{\infty}^{t} \frac{d\varepsilon_{\mathrm{d},k}(t^{\prime})}{dt^{\prime}}e^{-i(\omega_{\mathrm{d},k}-\omega_{\mathrm{r},k}) t^{\prime}}  dt^{\prime} .
\end{align}

It is reasonable to say that $\varepsilon_{\mathrm{d},k}(-\infty) = \alpha(-\infty) = 0$. Moreover, assuming the amplitude $\varepsilon_{\mathrm{d},k}(t)$ as a $C^{\infty}$-function (infinitely differentiable and analytical), we apply the Riemann--Lebesgue lemma~\cite{Serov:17} to the integral in the above equation and put it approximately to zero, one gets
\begin{align}
	\int_{-\infty}^{t} \varepsilon_{\mathrm{d},k}(t^{\prime})e^{-i(\omega_{\mathrm{d},k}-\omega_{\mathrm{r},k}) t^{\prime}} dt^{\prime} \approx 
	\frac{\varepsilon_{\mathrm{d},k}(t)e^{-i(\omega_{\mathrm{d},k}-\omega_{\mathrm{r},k}) t}}{-i(\omega_{\mathrm{d},k}-\omega_{\mathrm{r},k})} .
\end{align}

And therefore
\begin{align}
	\xi(t) = \frac{\varepsilon_{\mathrm{d},k}(t)}{\omega_{\mathrm{r},k}-\omega_{\mathrm{d},k}}e^{-i\omega_{\mathrm{d},k}t + i\phi_{\mathrm{d},k}} .
\end{align}

Using this result in the Eq.~\eqref{Ap-Eq:EffDrive}
\begin{align}
\hat{H}_{\mathrm{dr}}^{\mathrm{cl}}(t) &= \hbar \sum_{k=1}^{2}\sum_{n=1}^{N} \Omega_{k,n}(t) \left(\hat{a}_{k}^{\dagger} e^{-i\omega_{\mathrm{d},n}t + i\phi_{\mathrm{d},n}}  + \hat{a}_{k} e^{i\omega_{\mathrm{d},n}t - i\phi_{\mathrm{d},n}} \right) , \label{ApEq:HdrCl}
\end{align}
where $\Omega_{k,n}(t) = - g_{k}\varepsilon_{\mathrm{d},n}(t) / \Delta_{R,n}$, with $\Delta_{R,n}=\omega_{\mathrm{r},n}-\omega_{\mathrm{d},n}$, concluding then the demonstration of the final Hamiltonian for the drive given by the Eq.~\eqref{Eq:HdrCl} of the main text.

\section{The Baker-Campbell-Hausdorff expansion} \label{ApSec:Effective}

Now, we describe how to ``remove" the coupling with the resonator by applying a transformation of the $\hat{R}(\eta) = \exp\!~(\hat{S}(\eta))$, with $\hat{S}(\eta)$ defined in Eq.~\eqref{Eq:STranf}. Now, let us find the parameters $\eta_{k}$'s in order to eliminate the atoms-resonators interactions in second order. To this end we use the Baker-Campbell-Hausdorff expansion up to second order as
\begin{align}
	\hat{R} \hat{H}_{\Dcal}(t)\hat{R}^{\dagger} = \hat{H}_{\Dcal}(t) + [\hat{S},\hat{H}_{\Dcal}(t)] + \frac{1}{2!}\left[\hat{S}, [\hat{S},\hat{H}_{\Dcal}(t)] \right] . \label{Ap-Eq:BCHexp}
\end{align}

Here we are assuming that the drive (external field) is turned off, since we are interested only in the interaction between the atoms and the resonator. In this case, we have
\begin{align}
	\hat{R} \hat{H}_{\Dcal}(t)\hat{R}^{\dagger} &= \hat{H}_{\mathrm{int},0} + \hat{H}_{\alpha} + \hat{H}_{\mathrm{r}} + \hat{H}_{\mathrm{int}} + [\hat{S},\hat{H}_{\mathrm{int},0} + \hat{H}_{\alpha} + \hat{H}_{\mathrm{r}}] \nonumber \\
	&+ [\hat{S},\hat{H}_{\mathrm{int}}] +\frac{1}{2!}\left[\hat{S}, [\hat{S},\hat{H}_{\mathrm{int},0} + \hat{H}_{\alpha} + \hat{H}_{\mathrm{r}} + \hat{H}_{\mathrm{int}}] \right] . \label{Ap-Eq:BCHexp_2}
\end{align}

First, notice that
\begin{align}
	[\hat{S},\hat{H}_{\mathrm{int},0} + \hat{H}_{\mathrm{r}}] &= 
	\hbar \sum_{n=1,2} \omega_{n} \eta_{1} ([\hat{a}_{1}^{\dagger}\hat{r},\hat{a}_{n}^{\dagger}\hat{a}_{n}] - [\hat{a}_{1}\hat{r}^{\dagger} ,\hat{a}_{n}^{\dagger}\hat{a}_{n}])  \nonumber \\
	&
	+ \hbar \sum_{n=1,2} \omega_{n}\eta_{2} ([\hat{a}_{2}^{\dagger}\hat{r},\hat{a}_{n}^{\dagger}\hat{a}_{n}] - [\hat{a}_{2}\hat{r}^{\dagger},\hat{a}_{n}^{\dagger}\hat{a}_{n}])  \nonumber \\
	&+
	\eta_{1} \hbar g [(\hat{a}_{1}^{\dagger}\hat{r} - \hat{a}_{1}\hat{r}^{\dagger}),\hat{a}_{1}^{\dagger}\hat{a}_{2} + \hat{a}_{1}\hat{a}_{2}^{\dagger}] \nonumber \\
	&+ \hbar g \eta_{2} [(\hat{a}_{2}^{\dagger}\hat{r} - \hat{a}_{2}\hat{r}^{\dagger}) ,\hat{a}_{1}^{\dagger}\hat{a}_{2} + \hat{a}_{1}\hat{a}_{2}^{\dagger}]	,
\end{align}
and we use that $[\hat{a}_{k},\hat{a}_{n}^{\dagger}\hat{a}_{n}] = \delta_{kn}\hat{a}_{n}$ to get
\begin{align}
	[\hat{S},\hat{H}_{\mathrm{int},0} + \hat{H}_{\mathrm{r}}] 
	&=\hbar\left( \eta_{1} \Delta_{1} -
	g \eta_{2}\right) (\hat{a}_{1}^{\dagger}\hat{r} + \hat{a}_{1}\hat{r}^{\dagger}) 
	\nonumber \\
	&+
	\hbar \left(\eta_{2} \Delta_{2} - g \eta_{1} \right) (\hat{a}_{2}^{\dagger}\hat{r} + \hat{a}_{2}\hat{r}^{\dagger})
\end{align}
with $\Delta_{n} = \omega_{\mathrm{r}} - \omega_{n}$. To eliminate the interaction with the resonator we impose $[\hat{S},\hat{H}_{\mathrm{int},0} + \hat{H}_{\mathrm{r}}]=-\hat{H}_{\mathrm{int}}$, we need to satisfy the system of equations
$\eta_{1} \Delta_{1} - g \eta_{2} = -g_{1}$ and $\eta_{2} \Delta_{2} - g \eta_{1} = -g_{2}$, where the solution reads
\begin{align}
	\eta_{1} = \frac{g g_{2}+\Delta_{2} g_{1}}{g^2-\Delta_{1} \Delta_{2}} , \quad 
	\eta_{2} = \frac{g g_{1}+\Delta_{1} g_{2}}{g^2-\Delta_{1} \Delta_{2}} .
\end{align}

Then, using these choice in Eq.~\eqref{Ap-Eq:BCHexp_2}, we get
\begin{align}
	\hat{R} \hat{H}_{\Dcal}(t)\hat{R}^{\dagger} &= 
	\hat{H}_{\mathrm{qb}} + \hat{H}_{\mathrm{r}} + \frac{1}{2}[\hat{S}, \hat{H}_{\mathrm{int}}] 
	+ \left[\hat{S}, [\hat{S},\hat{H}_{\mathrm{int}}] \right] 
	\nonumber \\
	&+
	[\hat{S}, \hat{H}_{\alpha} ] + \frac{1}{2!}\left[\hat{S}, [\hat{S}, \hat{H}_{\alpha} ] \right] ,
\end{align}
where we can neglect the term $\left[\hat{S}, [\hat{S},\hat{H}_{\mathrm{int}}] \right]$ because it is proportional to terms of the order of $g_{k}^{2}/\Delta_{k}^{2}$. In conclusion, we get
\begin{align}
	\hat{R} \hat{H}_{\Dcal}(t)\hat{R}^{\dagger} = 
	\hat{H}_{\mathrm{qb}} + \hat{H}_{\mathrm{r}} + \frac{1}{2}[\hat{S}, \hat{H}_{\mathrm{int}}] + \hat{H}_{\mathrm{eff},\alpha}.
\end{align}
where $\hat{H}_{\mathrm{eff},\alpha}$ is the contribution to the effective Hamiltonian due to the anharmonicity of the system given by
\begin{align}
	\hat{H}_{\mathrm{eff},\alpha} = [\hat{S}, \hat{H}_{\alpha} ] + \frac{1}{2!}\left[\hat{S}, [\hat{S}, \hat{H}_{\alpha} ] \right].
\end{align}

We will discuss its influence when appropriate. By solving the commutator $[\hat{S}, \hat{H}_{\mathrm{int}}]$, we can write
\begin{align}
	[\hat{S}, \hat{H}_{\mathrm{int}}] 
	&=-2\left(\hbar g_{1}\eta_{1} + \hbar g_{2}\eta_{2}\right) \hat{r}^{\dagger} \hat{r} 
	+
	\hbar \sum_{n=1,2} 2g_{n}\eta_{n} 
	\hat{a}_{n}^{\dagger}\hat{a}_{n}
	\nonumber \\
	&+
	\hbar\left( g_{2}\eta_{1} + g_{1}\eta_{2} \right)\left(\hat{a}_{2}^{\dagger}\hat{a}_{1} +\hat{a}_{2}\hat{a}_{1}^{\dagger} \right).
\end{align}
where we neglected the zero-energy displacement proportional to identity. In this way, we use the Eq.~\eqref{Ap-Eq:BCHexp} and the effective Hamiltonian becomes
\begin{align}
	\hat{H}_{\mathrm{eff}}(t)&= \hbar \left[\omega_{\mathrm{r}} -\left(g_{1}\eta_{1} + g_{2}\eta_{2}\right)\right] \hat{r}^{\dagger}\hat{r}+
	\hbar g_{\mathrm{eff}} \left(\hat{a}_{1}^{\dagger}\hat{a}_{2} + \hat{a}_{1}\hat{a}_{2}^{\dagger}\right) \nonumber \\
	&+ \hbar \sum_{n=1,2} \left[\left(\omega_{n} + g_{n}\eta_{n}\right)\hat{a}_{n}^{\dagger}\hat{a}_{n} + \frac{\alpha_{n}}{2}\hat{a}_{n}^{\dagger}\hat{a}_{n}^{\dagger}\hat{a}_{n}\hat{a}_{n}\right] ,
\end{align}
with the total effective coupling given by 
\begin{align}
	g_{\mathrm{eff}} = g_{2}\eta_{1} + g_{1}\eta_{2} = g + \frac{g \left(g_{1}^2+g_{2}^2\right)+g_{1}g_{2} (\Delta_{1} +\Delta_{2} )}{2(g^2-\Delta_{1} \Delta_{2})}  . \label{Ap-Eq:g_eff}
\end{align}

\section{Contribution of the anharmonicity to the effective dynamics} \label{ApSec:Anharmonicity}

Let us consider the anharmonicity term in the BCH formula
\begin{align}
	\hat{H}_{\mathrm{eff},\alpha} = [\hat{S}, \hat{H}_{\alpha} ] + \frac{1}{2!}\left[\hat{S}, [\hat{S}, \hat{H}_{\alpha} ] \right] .
\end{align}

Again, the last term can be neglected in second order of $g_{k}/\Delta_{k}$, then we focus on the first term that reads
\begin{align}
	[\hat{S}, \hat{H}_{\alpha} ] 
	&= \hbar \frac{\alpha_{1}\eta_{1} }{2}\left\{
	\hat{r}\left[\hat{a}_{1}^{\dagger},\hat{a}_{1}^{\dagger}\hat{a}_{1}^{\dagger}\hat{a}_{1}\hat{a}_{1}\right]
	-
	\hat{r}^{\dagger}\left[\hat{a}_{1},\hat{a}_{1}^{\dagger}\hat{a}_{1}^{\dagger}\hat{a}_{1}\hat{a}_{1}\right]
	\right\}\nonumber 
	\\&
	+
	\hbar \frac{\alpha_{n}\eta_{2} }{2}\left\{
	\hat{r}\left[\hat{a}_{2}^{\dagger},\hat{a}_{2}^{\dagger}\hat{a}_{2}^{\dagger}\hat{a}_{2}\hat{a}_{2}\right]
	-
	\hat{r}^{\dagger}\left[\hat{a}_{2},\hat{a}_{2}^{\dagger}\hat{a}_{2}^{\dagger}\hat{a}_{2}\hat{a}_{2}\right]
	\right\}
\end{align}

Now, we use that $\left[\hat{a}_{n}^{\dagger},\hat{a}_{n}^{\dagger}\hat{a}_{n}^{\dagger}\hat{a}_{n}\hat{a}_{n}\right] = -2\hat{a}_{n}^{\dagger}\hat{a}_{n}^{\dagger}\hat{a}_{n}$ to get
\begin{align}
	[\hat{S}, \hat{H}_{\alpha} ]
	&= -\hbar \alpha_{1}\eta_{1}\left(
	\hat{a}_{1}^{\dagger}\hat{a}_{1}^{\dagger}\hat{a}_{1}\hat{r}	+	\hat{a}_{1}^{\dagger}\hat{a}_{1}\hat{a}_{1}\hat{r}^{\dagger}
	\right)\nonumber 
	\\&
	-
	\hbar \alpha_{2}\eta_{2} \left(
	\hat{a}_{2}^{\dagger}\hat{a}_{2}^{\dagger}\hat{a}_{2}\hat{r}	+	\hat{a}_{2}^{\dagger}\hat{a}_{2}\hat{a}_{2}\hat{r}^{\dagger}
	\right) .
\end{align}

Now, by applying the above Hamiltonian to a state of the form $\ket{\Psi} = \ket{\psi_{\mathrm{atoms}}} \otimes (a\ket{0} + b\ket{1})$, where the resonator is in the low excitation regime $|b|^2 \ll |a|^2$. We get
\begin{align}
	[\hat{S}, \hat{H}_{\alpha} ] \ket{\Psi} &= -\hbar b\left( \alpha_{1}\eta_{1}
	\hat{a}_{1}^{\dagger}\hat{a}_{1}^{\dagger}\hat{a}_{1} + \alpha_{2}\eta_{2}\hat{a}_{2}^{\dagger}\hat{a}_{2}^{\dagger}\hat{a}_{2}\right)\ket{\psi_{\mathrm{atoms}}}\ket{0}
	\nonumber 
	\\
	&
	-\hbar \alpha_{1}\eta_{1} \hat{a}_{1}^{\dagger}\hat{a}_{1}\hat{a}_{1}\ket{\psi_{\mathrm{atoms}}}(a\ket{1} + b\sqrt{2}\ket{2}) 
	\nonumber 
	\\
	&  -	\hbar \alpha_{2}\eta_{2}	\hat{a}_{2}^{\dagger}\hat{a}_{2}\hat{a}_{2}\ket{\psi_{\mathrm{atoms}}}(a\ket{1} + b\sqrt{2}\ket{2}) .
\end{align}

Now, we can analyze the effects of such a term in the system. The above equation suggests that there is a \textit{low} probability $|b|^2$ of the resonator losing one excitation to the atom, since the operation $\hat{a}_{n}^{\dagger}\hat{a}_{n}^{\dagger}\hat{a}_{n}\ket{\psi_{\mathrm{atoms}}}$ creates an excitation in the system. On the other hand, the terms proportional to $|a|^2$ state that a double excitation cascade in the atoms can occur, followed by an excitation. While such a process can be possible when each atom is doubly excited, the probability of such a process in the qubit subspace is zero and such a term can be neglected in these cases. In conclusion, it is reasonable to assume $[\hat{S}, \hat{H}_{\alpha} ] \ket{\Psi} \approx 0$ in our study.

\section{The Landau-Zener like Hamiltonian} \label{ApSec:LandauZener}

Consider the following Hamiltonian
\begin{align}
	\hat{H}_{\mathrm{qubit}} &= \hbar \left(\tilde{\omega}_{1}\hat{\sigma}_{1}^{+}\hat{\sigma}_{1}^{-} + \tilde{\omega}_{2}\hat{\sigma}_{2}^{+}\hat{\sigma}_{2}^{-}\right) +
	\hbar g_{\mathrm{eff}} \left(\hat{\sigma}_{1}^{+}\hat{\sigma}_{2}^{-} + \hat{\sigma}_{1}^{-}\hat{\sigma}_{2}^{+}\right) \nonumber \\
	&+\hbar \sum\nolimits_{k=1}^{2}\Omega_{k}(t) \left(\hat{\sigma}_{k}^{+} e^{-i\omega_{\mathrm{d}}t + i\phi_{\mathrm{d}}}  + \hat{\sigma}_{k}^{-} e^{i\omega_{\mathrm{d}}t - i\phi_{\mathrm{d}}} \right) , \label{ApEq:H_qubit}
\end{align}
obtained from Eq.~\eqref{Eq:H_qubit} for the case of a single drive with frequency $\omega_{\mathrm{d}}$ and phase $\phi_{\mathrm{d}}$. The approach used in Appendix~\ref{ApSec:SemiClass} allows us to rewrite the Schrödinger equation in a new frame given by an arbitrary unitary operator $\hat{R}(t)$, with the new Hamiltonian
\begin{align}
	\hat{H}_{R}(t) = \hat{R}(t)\hat{H}_{\mathrm{qubit}}\hat{R}^{\dagger}(t) + i\hbar\frac{d\hat{R}(t)}{dt} \hat{R}^{\dagger}(t) .
\end{align}

The Hamiltonian in Eq.~\eqref{Eq:H_prime} is obtained from the operator
\begin{align}
\hat{R}(t) = \exp \left( i \omega_{\mathrm{d}}t\hat{\sigma}_{1}^{+}\hat{\sigma}_{1}^{-} + i \omega_{\mathrm{d}}t\hat{\sigma}_{2}^{+}\hat{\sigma}_{2}^{-} \right) .
\end{align}

In fact, by using that $\hat{R}(t)\hat{\sigma}_{k}^{\pm}\hat{R}^{\dagger}(t) = e^{\pm i \omega_{\mathrm{d}}t}\hat{\sigma}_{k}^{\pm}$ and
\begin{align}
i\hbar\frac{d\hat{R}(t)}{dt} \hat{R}^{\dagger}(t) &= i\hbar\left[i \omega_{\mathrm{d}}\hat{\sigma}_{1}^{+}\hat{\sigma}_{1}^{-} + i \omega_{\mathrm{d}}\hat{\sigma}_{2}^{+}\hat{\sigma}_{2}^{-}\right]\hat{R}(t)\hat{R}^{\dagger}(t) \nonumber \\
&= - \hbar\omega_{\mathrm{d}}\left[\hat{\sigma}_{1}^{+}\hat{\sigma}_{1}^{-} + \hat{\sigma}_{2}^{+}\hat{\sigma}_{2}^{-}\right],
\end{align}
where we used $\hat{R}(t)\hat{R}^{\dagger}(t)=\1$, one finds
\begin{align}
\hat{H}_{R}(t) &= \hbar \left(\tilde{\omega}_{1}\hat{\sigma}_{1}^{+}\hat{\sigma}_{1}^{-} + \tilde{\omega}_{2}\hat{\sigma}_{2}^{+}\hat{\sigma}_{2}^{-}\right) +
\hbar g_{\mathrm{eff}} \left(\hat{\sigma}_{1}^{+}\hat{\sigma}_{2}^{-} + \hat{\sigma}_{1}^{-}\hat{\sigma}_{2}^{+}\right) \nonumber \\
&+\hbar \sum\nolimits_{k=1}^{2}\Omega_{k}(t) \left(\hat{\sigma}_{k}^{+} e^{i\phi_{\mathrm{d}}}  + \hat{\sigma}_{k}^{-} e^{- i\phi_{\mathrm{d}}} \right) \nonumber \\
&- \hbar\omega_{\mathrm{d}}\left[\hat{\sigma}_{1}^{+}\hat{\sigma}_{1}^{-} + \hat{\sigma}_{2}^{+}\hat{\sigma}_{2}^{-}\right]
\end{align}
or similarly
\begin{align}
	\hat{H}_{R}(t) &= \hbar \sum\nolimits_{k=1}^{2}\left[\left(\tilde{\omega}_{1}-\omega_{\mathrm{d}}\right)\hat{\sigma}_{1}^{+}\hat{\sigma}_{1}^{-} + \Omega_{k}(t) \left(\hat{\sigma}_{k}^{+} e^{i\phi_{\mathrm{d}}}  + \hat{\sigma}_{k}^{-} e^{- i\phi_{\mathrm{d}}} \right)\right] \nonumber \\
	&+
	\hbar g_{\mathrm{eff}} \left(\hat{\sigma}_{1}^{+}\hat{\sigma}_{2}^{-} + \hat{\sigma}_{1}^{-}\hat{\sigma}_{2}^{+}\right) .
\end{align}

Therefore, it concludes that $\hat{H}_{R}(t) = \hat{H}^{\prime}$, with $\hat{H}^{\prime}$ given by Eq.~\eqref{Eq:H_prime}.

\begin{figure}[t!]
	\includegraphics[width=\columnwidth]{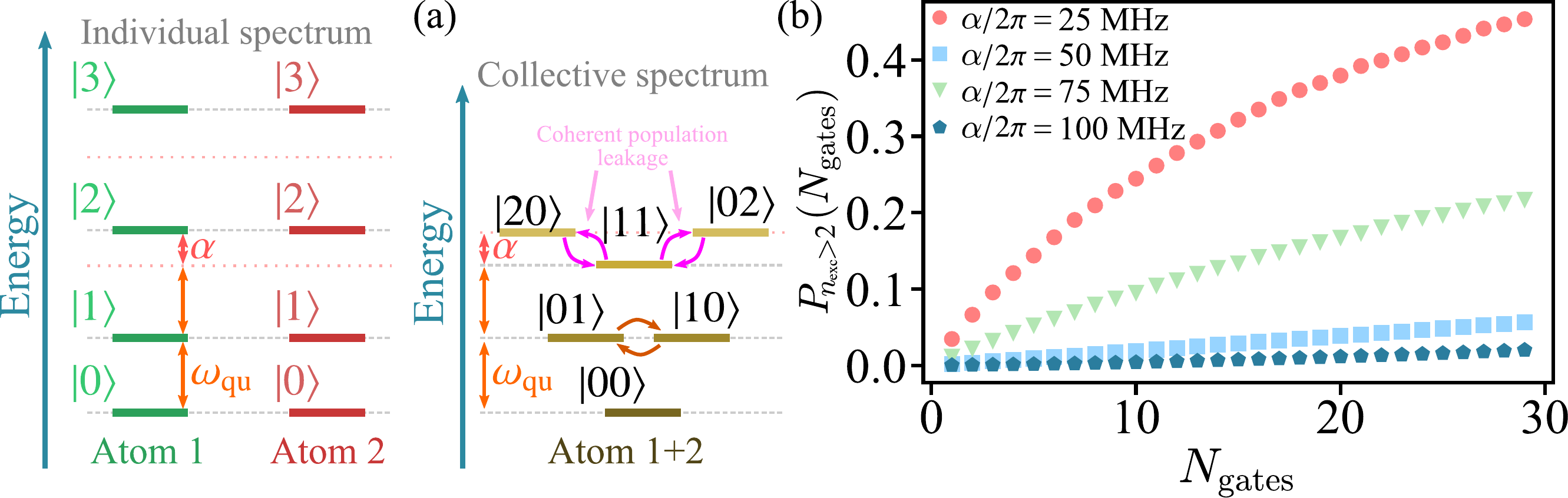}
	\caption{(a) Individual and collective spectrum of two identical superconducting atoms with frequency transition $\omega_{\mathrm{qu}}$ and anharmonicity $\alpha$. The representation of the collective spectrum illustrates the proximity (in energy) between collective states of the doubly excited subspace. Quantum states of sectors with different number of excitation are separated by the bare energy of the qubit $\hbar \omega_{\mathrm{qu}}$, while energy splitting between states of a given sector is possible due to a non-zero anharmonicity $\alpha$. For the case of system with weak anharmonicity transitions are allowed between the states $\ket{11}$, $\ket{20}$ and $\ket{02}$ (magenta arrows).
		(b) Population leakage as a function of the number of quantum gates for different values of anharmonicity. To this graph we use the same parameters as in Fig.~\ref{Fig:MatrixProcess}, for the case where parasitic interaction is taken into account.}\label{AP-Fig:PopulationLeackage}
\end{figure}

\section{Population leakage in weakly anharmonic atoms} \label{ApSec:WeakAnharmonicity}

In this section we briefly show how population leakage can affect the circuit implementation shown in Fig.~\ref{Fig:MatrixProcess} of the main text. As shown in Fig.~\SubFig{AP-Fig:PopulationLeackage}{a}, the qubit subspace of a two-qubit system is $\{\ket{00},\ket{01},\ket{10},\ket{11}\}$, where population in a highly excited state is avoided whenever it is possible. However, in weakly anharmonic atoms, the energy level splitting between the states of the doubly excited state sector become comparable to the atom-atom interaction. In this scenario, non-negligible transitions $\ket{11} \rightleftarrows \ket{02}\rightleftarrows \ket{20}$ can lead to a coherent ``population leakage" from the qubit subspace to highly excited states. To quantify such a population leakage, let us define the figure of merit as
\begin{align}
P_{n_{\mathrm{exc}}>2}(N_{\mathrm{gates}}) = \trace \left[ (\1 - \hat{P}_{\mathrm{qu}}) \rho(N_{\mathrm{gates}})\right] ,
\end{align}
with $\hat{P}_{\mathrm{qu}} = \sum_{n=0,m=0}^{1,1}\ket{nm}\bra{nm}$ the projector defined over the qubit subspace of the system. In this way, $\1 - \hat{P}_{\mathrm{qu}}$ is a projector outside the qubit subspace and $P_{n_{\mathrm{exc}}>2}(N_{\mathrm{gates}})$ quantifies the population of the subspace with total number of excitation $n_{\mathrm{exc}} > 2$. In the case of two ideal qubits we should have $P_{n_{\mathrm{exc}}>2}(N_{\mathrm{gates}}) = 0$. By assuming two qubits with identical anharmonicity $\alpha_{1}=\alpha_{2}=\alpha$ and frequencies $\omega_{1}=\omega_{2}=\omega_{\mathrm{qu}}$, in Fig.~\SubFig{AP-Fig:PopulationLeackage}{b} we show the behavior of the population $P_{n_{\mathrm{exc}}>2}(N_{\mathrm{gates}})$ for different values of anharmonicity $\alpha$ as a function of the number of $i$SWAP gates $N_{\mathrm{gates}}$ for the circuit considered in Fig.~\ref{Fig:MatrixProcess}. In conclusion, even for a (relatively) small circuit ($30$ gates) we have a strong population leakage when anharmonicity is not large enough.

\newpage


\begin{thebibliography}{59}%
	\makeatletter
	\providecommand \@ifxundefined [1]{%
		\@ifx{#1\undefined}
	}%
	\providecommand \@ifnum [1]{%
		\ifnum #1\expandafter \@firstoftwo
		\else \expandafter \@secondoftwo
		\fi
	}%
	\providecommand \@ifx [1]{%
		\ifx #1\expandafter \@firstoftwo
		\else \expandafter \@secondoftwo
		\fi
	}%
	\providecommand \natexlab [1]{#1}%
	\providecommand \enquote  [1]{``#1''}%
	\providecommand \bibnamefont  [1]{#1}%
	\providecommand \bibfnamefont [1]{#1}%
	\providecommand \citenamefont [1]{#1}%
	\providecommand \href@noop [0]{\@secondoftwo}%
	\providecommand \href [0]{\begingroup \@sanitize@url \@href}%
	\providecommand \@href[1]{\@@startlink{#1}\@@href}%
	\providecommand \@@href[1]{\endgroup#1\@@endlink}%
	\providecommand \@sanitize@url [0]{\catcode `\\12\catcode `\$12\catcode
		`\&12\catcode `\#12\catcode `\^12\catcode `\_12\catcode `\%12\relax}%
	\providecommand \@@startlink[1]{}%
	\providecommand \@@endlink[0]{}%
	\providecommand \url  [0]{\begingroup\@sanitize@url \@url }%
	\providecommand \@url [1]{\endgroup\@href {#1}{\urlprefix }}%
	\providecommand \urlprefix  [0]{URL }%
	\providecommand \Eprint [0]{\href }%
	\providecommand \doibase [0]{http://dx.doi.org/}%
	\providecommand \selectlanguage [0]{\@gobble}%
	\providecommand \bibinfo  [0]{\@secondoftwo}%
	\providecommand \bibfield  [0]{\@secondoftwo}%
	\providecommand \translation [1]{[#1]}%
	\providecommand \BibitemOpen [0]{}%
	\providecommand \bibitemStop [0]{}%
	\providecommand \bibitemNoStop [0]{.\EOS\space}%
	\providecommand \EOS [0]{\spacefactor3000\relax}%
	\providecommand \BibitemShut  [1]{\csname bibitem#1\endcsname}%
	\let\auto@bib@innerbib\@empty
	\bibitem [{\citenamefont {Song}\ \textit {et~al.}(2017)\citenamefont {Song},
		\citenamefont {Xu}, \citenamefont {Liu}, \citenamefont {Yang}, \citenamefont
		{Zheng}, \citenamefont {Deng}, \citenamefont {Xie}, \citenamefont {Huang},
		\citenamefont {Guo}, \citenamefont {Zhang}, \citenamefont {Zhang},
		\citenamefont {Xu}, \citenamefont {Zheng}, \citenamefont {Zhu}, \citenamefont
		{Wang}, \citenamefont {Chen}, \citenamefont {Lu}, \citenamefont {Han},\ and\
		\citenamefont {Pan}}]{Song:17PRL}%
	\BibitemOpen
	\bibfield  {author} {\bibinfo {author} {\bibfnamefont {C.}~\bibnamefont
			{Song}},  {\em et~al.},\ }\enquote{\bibinfo {title} {10-Qubit Entanglement
			and Parallel Logic Operations with a Superconducting Circuit}},\ \href
	{\doibase 10.1103/PhysRevLett.119.180511} {\bibfield  {journal} {\bibinfo
			{journal} {Phys. Rev. Lett.}\ }\textbf {\bibinfo {volume} {119}},\ \bibinfo
		{pages} {180511} (\bibinfo {year} {2017})}\BibitemShut {NoStop}%
	\bibitem [{\citenamefont {Gong}\ \textit {et~al.}(2019)\citenamefont {Gong},
		\citenamefont {Chen}, \citenamefont {Zheng}, \citenamefont {Wang},
		\citenamefont {Zha}, \citenamefont {Deng}, \citenamefont {Yan}, \citenamefont
		{Rong}, \citenamefont {Wu}, \citenamefont {Li}, \citenamefont {Chen},
		\citenamefont {Zhao}, \citenamefont {Liang}, \citenamefont {Lin},
		\citenamefont {Xu}, \citenamefont {Guo}, \citenamefont {Sun}, \citenamefont
		{Castellano}, \citenamefont {Wang}, \citenamefont {Peng}, \citenamefont {Lu},
		\citenamefont {Zhu},\ and\ \citenamefont {Pan}}]{Gong:19}%
	\BibitemOpen
	\bibfield  {author} {\bibinfo {author} {\bibfnamefont {M.}~\bibnamefont
			{Gong}},  {\em et~al.},\ }\enquote{\bibinfo {title} {Genuine 12-Qubit
			Entanglement on a Superconducting Quantum Processor}},\ \href {\doibase
		10.1103/PhysRevLett.122.110501} {\bibfield  {journal} {\bibinfo  {journal}
			{Phys. Rev. Lett.}\ }\textbf {\bibinfo {volume} {122}},\ \bibinfo {pages}
		{110501} (\bibinfo {year} {2019})}\BibitemShut {NoStop}%
	\bibitem [{\citenamefont {Arute}\ \textit {et~al.}(2019)\citenamefont {Arute},
		\citenamefont {Arya}, \citenamefont {Babbush}, \citenamefont {Bacon},
		\citenamefont {Bardin}, \citenamefont {Barends}, \citenamefont {Biswas},
		\citenamefont {Boixo}, \citenamefont {Brandao}, \citenamefont {Buell},
		\citenamefont {Burkett}, \citenamefont {Chen}, \citenamefont {Chen},
		\citenamefont {Chiaro}, \citenamefont {Collins}, \citenamefont {Courtney},
		\citenamefont {Dunsworth}, \citenamefont {Farhi}, \citenamefont {Foxen},
		\citenamefont {Fowler}, \citenamefont {Gidney}, \citenamefont {Giustina},
		\citenamefont {Graff}, \citenamefont {Guerin}, \citenamefont {Habegger},
		\citenamefont {Harrigan}, \citenamefont {Hartmann}, \citenamefont {Ho},
		\citenamefont {Hoffmann}, \citenamefont {Huang}, \citenamefont {Humble},
		\citenamefont {Isakov}, \citenamefont {Jeffrey}, \citenamefont {Jiang},
		\citenamefont {Kafri}, \citenamefont {Kechedzhi}, \citenamefont {Kelly},
		\citenamefont {Klimov}, \citenamefont {Knysh}, \citenamefont {Korotkov},
		\citenamefont {Kostritsa}, \citenamefont {Landhuis}, \citenamefont
		{Lindmark}, \citenamefont {Lucero}, \citenamefont {Lyakh}, \citenamefont
		{Mandrà}, \citenamefont {McClean}, \citenamefont {McEwen}, \citenamefont
		{Megrant}, \citenamefont {Mi}, \citenamefont {Michielsen}, \citenamefont
		{Mohseni}, \citenamefont {Mutus}, \citenamefont {Naaman}, \citenamefont
		{Neeley}, \citenamefont {Neill}, \citenamefont {Niu}, \citenamefont {Ostby},
		\citenamefont {Petukhov}, \citenamefont {Platt}, \citenamefont {Quintana},
		\citenamefont {Rieffel}, \citenamefont {Roushan}, \citenamefont {Rubin},
		\citenamefont {Sank}, \citenamefont {Satzinger}, \citenamefont {Smelyanskiy},
		\citenamefont {Sung}, \citenamefont {Trevithick}, \citenamefont
		{Vainsencher}, \citenamefont {Villalonga}, \citenamefont {White},
		\citenamefont {Yao}, \citenamefont {Yeh}, \citenamefont {Zalcman},
		\citenamefont {Neven},\ and\ \citenamefont {Martinis}}]{Arute:19}%
	\BibitemOpen
	\bibfield  {author} {\bibinfo {author} {\bibfnamefont {F.}~\bibnamefont
			{Arute}},  {\em et~al.},\ }\enquote{\bibinfo {title} {Quantum supremacy using
			a programmable superconducting processor}},\ \href {\doibase
		10.1038/s41586-019-1666-5} {\bibfield  {journal} {\bibinfo  {journal}
			{Nature}\ }\textbf {\bibinfo {volume} {574}},\ \bibinfo {pages} {505}
		(\bibinfo {year} {2019})}\BibitemShut {NoStop}%
	\bibitem [{\citenamefont {Wu}\ \textit {et~al.}(2021)\citenamefont {Wu},
		\citenamefont {Bao}, \citenamefont {Cao}, \citenamefont {Chen}, \citenamefont
		{Chen}, \citenamefont {Chen}, \citenamefont {Chung}, \citenamefont {Deng},
		\citenamefont {Du}, \citenamefont {Fan}, \citenamefont {Gong}, \citenamefont
		{Guo}, \citenamefont {Guo}, \citenamefont {Guo}, \citenamefont {Han},
		\citenamefont {Hong}, \citenamefont {Huang}, \citenamefont {Huo},
		\citenamefont {Li}, \citenamefont {Li}, \citenamefont {Li}, \citenamefont
		{Li}, \citenamefont {Liang}, \citenamefont {Lin}, \citenamefont {Lin},
		\citenamefont {Qian}, \citenamefont {Qiao}, \citenamefont {Rong},
		\citenamefont {Su}, \citenamefont {Sun}, \citenamefont {Wang}, \citenamefont
		{Wang}, \citenamefont {Wu}, \citenamefont {Xu}, \citenamefont {Yan},
		\citenamefont {Yang}, \citenamefont {Yang}, \citenamefont {Ye}, \citenamefont
		{Yin}, \citenamefont {Ying}, \citenamefont {Yu}, \citenamefont {Zha},
		\citenamefont {Zhang}, \citenamefont {Zhang}, \citenamefont {Zhang},
		\citenamefont {Zhang}, \citenamefont {Zhao}, \citenamefont {Zhao},
		\citenamefont {Zhou}, \citenamefont {Zhu}, \citenamefont {Lu}, \citenamefont
		{Peng}, \citenamefont {Zhu},\ and\ \citenamefont {Pan}}]{Wu:21}%
	\BibitemOpen
	\bibfield  {author} {\bibinfo {author} {\bibfnamefont {Y.}~\bibnamefont
			{Wu}},  {\em et~al.},\ }\enquote{\bibinfo {title} {Strong Quantum
			Computational Advantage Using a Superconducting Quantum Processor}},\ \href
	{\doibase 10.1103/PhysRevLett.127.180501} {\bibfield  {journal} {\bibinfo
			{journal} {Phys. Rev. Lett.}\ }\textbf {\bibinfo {volume} {127}},\ \bibinfo
		{pages} {180501} (\bibinfo {year} {2021})}\BibitemShut {NoStop}%
	\bibitem [{\citenamefont {Gong}\ \textit {et~al.}(2021)\citenamefont {Gong},
		\citenamefont {de~Moraes~Neto}, \citenamefont {Zha}, \citenamefont {Wu},
		\citenamefont {Rong}, \citenamefont {Ye}, \citenamefont {Li}, \citenamefont
		{Zhu}, \citenamefont {Wang}, \citenamefont {Zhao}, \citenamefont {Liang},
		\citenamefont {Lin}, \citenamefont {Xu}, \citenamefont {Peng}, \citenamefont
		{Deng}, \citenamefont {Bayat}, \citenamefont {Zhu},\ and\ \citenamefont
		{Pan}}]{Gong:21}%
	\BibitemOpen
	\bibfield  {author} {\bibinfo {author} {\bibfnamefont {M.}~\bibnamefont
			{Gong}},  {\em et~al.},\ }\enquote{\bibinfo {title} {Experimental
			characterization of the quantum many-body localization transition}},\ \href
	{\doibase 10.1103/PhysRevResearch.3.033043} {\bibfield  {journal} {\bibinfo
			{journal} {Phys. Rev. Research}\ }\textbf {\bibinfo {volume} {3}},\ \bibinfo
		{pages} {033043} (\bibinfo {year} {2021})}\BibitemShut {NoStop}%
	\bibitem [{\citenamefont {Wendin}(2017)}]{Wendin:17}%
	\BibitemOpen
	\bibfield  {author} {\bibinfo {author} {\bibfnamefont {G.}~\bibnamefont
			{Wendin}},\ }\enquote{\bibinfo {title} {Quantum information processing with
			superconducting circuits: a review}},\ \href {\doibase
		10.1088/1361-6633/aa7e1a} {\bibfield  {journal} {\bibinfo  {journal} {Reports
				on Progress in Physics}\ }\textbf {\bibinfo {volume} {80}},\ \bibinfo {pages}
		{106001} (\bibinfo {year} {2017})}\BibitemShut {NoStop}%
	\bibitem [{\citenamefont {Siddiqi}(2021)}]{Siddiqi:21}%
	\BibitemOpen
	\bibfield  {author} {\bibinfo {author} {\bibfnamefont {I.}~\bibnamefont
			{Siddiqi}},\ }\enquote{\bibinfo {title} {Engineering high-coherence
			superconducting qubits}},\ \href {\doibase
		https://doi.org/10.1038/s41578-021-00370-4} {\bibfield  {journal} {\bibinfo
			{journal} {Nature Reviews Materials}\ }\textbf {\bibinfo {volume} {6}},\
		\bibinfo {pages} {875} (\bibinfo {year} {2021})}\BibitemShut {NoStop}%
	\bibitem [{\citenamefont {Burnett}\ \textit {et~al.}(2019)\citenamefont
		{Burnett}, \citenamefont {Bengtsson}, \citenamefont {Scigliuzzo},
		\citenamefont {Niepce}, \citenamefont {Kudra}, \citenamefont {Delsing},\ and\
		\citenamefont {Bylander}}]{Burnett:19}%
	\BibitemOpen
	\bibfield  {author} {\bibinfo {author} {\bibfnamefont {J.~J.}\ \bibnamefont
			{Burnett}},  {\em et~al.},\ }\enquote{\bibinfo {title} {Decoherence
			benchmarking of superconducting qubits}},\ \href@noop {} {\bibfield
		{journal} {\bibinfo  {journal} {npj Quantum Information}\ }\textbf {\bibinfo
			{volume} {5}},\ \bibinfo {pages} {1} (\bibinfo {year} {2019})}\BibitemShut
	{NoStop}%
	\bibitem [{\citenamefont {Zhao}\ \textit
		{et~al.}(2022{\natexlab{a}})\citenamefont {Zhao}, \citenamefont {Linghu},
		\citenamefont {Li}, \citenamefont {Xu}, \citenamefont {Wang}, \citenamefont
		{Xue}, \citenamefont {Jin},\ and\ \citenamefont {Yu}}]{ZhaoPeng:22}%
	\BibitemOpen
	\bibfield  {author} {\bibinfo {author} {\bibfnamefont {P.}~\bibnamefont
			{Zhao}},  {\em et~al.},\ }\enquote{\bibinfo {title} {Quantum Crosstalk
			Analysis for Simultaneous Gate Operations on Superconducting Qubits}},\ \href
	{\doibase 10.1103/PRXQuantum.3.020301} {\bibfield  {journal} {\bibinfo
			{journal} {PRX Quantum}\ }\textbf {\bibinfo {volume} {3}},\ \bibinfo {pages}
		{020301} (\bibinfo {year} {2022}{\natexlab{a}})}\BibitemShut {NoStop}%
	\bibitem [{\citenamefont {Willsch}\ \textit {et~al.}(2017)\citenamefont
		{Willsch}, \citenamefont {Nocon}, \citenamefont {Jin}, \citenamefont
		{De~Raedt},\ and\ \citenamefont {Michielsen}}]{Willsch:17}%
	\BibitemOpen
	\bibfield  {author} {\bibinfo {author} {\bibfnamefont {D.}~\bibnamefont
			{Willsch}},  {\em et~al.},\ }\enquote{\bibinfo {title} {Gate-error analysis
			in simulations of quantum computers with transmon qubits}},\ \href {\doibase
		10.1103/PhysRevA.96.062302} {\bibfield  {journal} {\bibinfo  {journal} {Phys.
				Rev. A}\ }\textbf {\bibinfo {volume} {96}},\ \bibinfo {pages} {062302}
		(\bibinfo {year} {2017})}\BibitemShut {NoStop}%
	\bibitem [{\citenamefont {Krinner}\ \textit {et~al.}(2020)\citenamefont
		{Krinner}, \citenamefont {Lazar}, \citenamefont {Remm}, \citenamefont
		{Andersen}, \citenamefont {Lacroix}, \citenamefont {Norris}, \citenamefont
		{Hellings}, \citenamefont {Gabureac}, \citenamefont {Eichler},\ and\
		\citenamefont {Wallraff}}]{Krinner:20}%
	\BibitemOpen
	\bibfield  {author} {\bibinfo {author} {\bibfnamefont {S.}~\bibnamefont
			{Krinner}},  {\em et~al.},\ }\enquote{\bibinfo {title} {Benchmarking Coherent
			Errors in Controlled-Phase Gates due to Spectator Qubits}},\ \href {\doibase
		10.1103/PhysRevApplied.14.024042} {\bibfield  {journal} {\bibinfo  {journal}
			{Phys. Rev. Applied}\ }\textbf {\bibinfo {volume} {14}},\ \bibinfo {pages}
		{024042} (\bibinfo {year} {2020})}\BibitemShut {NoStop}%
	\bibitem [{\citenamefont {Han}\ \textit {et~al.}(2020)\citenamefont {Han},
		\citenamefont {Cai}, \citenamefont {Li}, \citenamefont {Wu}, \citenamefont
		{Ma}, \citenamefont {Ma}, \citenamefont {Wang}, \citenamefont {Zhang},
		\citenamefont {Song},\ and\ \citenamefont {Duan}}]{Han:20}%
	\BibitemOpen
	\bibfield  {author} {\bibinfo {author} {\bibfnamefont {X.~Y.}\ \bibnamefont
			{Han}},  {\em et~al.},\ }\enquote{\bibinfo {title} {Error analysis in
			suppression of unwanted qubit interactions for a parametric gate in a tunable
			superconducting circuit}},\ \href {\doibase 10.1103/PhysRevA.102.022619}
	{\bibfield  {journal} {\bibinfo  {journal} {Phys. Rev. A}\ }\textbf {\bibinfo
			{volume} {102}},\ \bibinfo {pages} {022619} (\bibinfo {year}
		{2020})}\BibitemShut {NoStop}%
	\bibitem [{\citenamefont {{Dai}}\ \textit {et~al.}(2022)\citenamefont {{Dai}},
		\citenamefont {{Trappen}}, \citenamefont {{Yang}}, \citenamefont
		{{Disseler}}, \citenamefont {{Basham}}, \citenamefont {{Gibson}},
		\citenamefont {{Melville}}, \citenamefont {{Niedzielski}}, \citenamefont
		{{Das}}, \citenamefont {{Kim}}, \citenamefont {{Yoder}}, \citenamefont
		{{Weber}}, \citenamefont {{Hirjibehedin}}, \citenamefont {{Lidar}},\ and\
		\citenamefont {{Lupascu}}}]{Dai:22}%
	\BibitemOpen
	\bibfield  {author} {\bibinfo {author} {\bibfnamefont {X.}~\bibnamefont
			{{Dai}}},  {\em et~al.},\ }\enquote{\bibinfo {title} {{Optimizing for
				periodicity: a model-independent approach to flux crosstalk calibration for
				superconducting circuits}}},\ \href@noop {} {\bibfield  {journal} {\bibinfo
			{journal} {arXiv e-prints}\ ,\ \bibinfo {eid} {arXiv:2211.01497}} (\bibinfo
		{year} {2022})},\ \Eprint {http://arxiv.org/abs/2211.01497} {arXiv:2211.01497
		[quant-ph]} \BibitemShut {NoStop}%
	\bibitem [{\citenamefont {C{\'o}rcoles}\ \textit {et~al.}(2015)\citenamefont
		{C{\'o}rcoles}, \citenamefont {Magesan}, \citenamefont {Srinivasan},
		\citenamefont {Cross}, \citenamefont {Steffen}, \citenamefont {Gambetta},\
		and\ \citenamefont {Chow}}]{Corcoles:15}%
	\BibitemOpen
	\bibfield  {author} {\bibinfo {author} {\bibfnamefont {A.~D.}\ \bibnamefont
			{C{\'o}rcoles}},  {\em et~al.},\ }\enquote{\bibinfo {title} {Demonstration of
			a quantum error detection code using a square lattice of four superconducting
			qubits}},\ \href {\doibase https://doi.org/10.1038/ncomms7979} {\bibfield
		{journal} {\bibinfo  {journal} {Nature communications}\ }\textbf {\bibinfo
			{volume} {6}},\ \bibinfo {pages} {1} (\bibinfo {year} {2015})}\BibitemShut
	{NoStop}%
	\bibitem [{\citenamefont {Tripathi}\ \textit {et~al.}(2022)\citenamefont
		{Tripathi}, \citenamefont {Chen}, \citenamefont {Khezri}, \citenamefont
		{Yip}, \citenamefont {Levenson-Falk},\ and\ \citenamefont
		{Lidar}}]{Tripathi:22}%
	\BibitemOpen
	\bibfield  {author} {\bibinfo {author} {\bibfnamefont {V.}~\bibnamefont
			{Tripathi}},  {\em et~al.},\ }\enquote{\bibinfo {title} {Suppression of
			Crosstalk in Superconducting Qubits Using Dynamical Decoupling}},\ \href
	{\doibase 10.1103/PhysRevApplied.18.024068} {\bibfield  {journal} {\bibinfo
			{journal} {Phys. Rev. Applied}\ }\textbf {\bibinfo {volume} {18}},\ \bibinfo
		{pages} {024068} (\bibinfo {year} {2022})}\BibitemShut {NoStop}%
	\bibitem [{\citenamefont {{Chen}}\ \textit {et~al.}(2021)\citenamefont
		{{Chen}}, \citenamefont {{Satzinger}}, \citenamefont {{Atalaya}},
		\citenamefont {{Korotkov}}, \citenamefont {{Dunsworth}}, \citenamefont
		{{Sank}}, \citenamefont {{Quintana}}, \citenamefont {{McEwen}}, \citenamefont
		{{Barends}}, \citenamefont {{Klimov}}, \citenamefont {{Hong}}, \citenamefont
		{{Jones}}, \citenamefont {{Petukhov}}, \citenamefont {{Kafri}}, \citenamefont
		{{Demura}}, \citenamefont {{Burkett}}, \citenamefont {{Gidney}},
		\citenamefont {{Fowler}}, \citenamefont {{Putterman}}, \citenamefont
		{{Aleiner}}, \citenamefont {{Arute}}, \citenamefont {{Arya}}, \citenamefont
		{{Babbush}}, \citenamefont {{Bardin}}, \citenamefont {{Bengtsson}},
		\citenamefont {{Bourassa}}, \citenamefont {{Broughton}}, \citenamefont
		{{Buckley}}, \citenamefont {{Buell}}, \citenamefont {{Bushnell}},
		\citenamefont {{Chiaro}}, \citenamefont {{Collins}}, \citenamefont
		{{Courtney}}, \citenamefont {{Derk}}, \citenamefont {{Eppens}}, \citenamefont
		{{Erickson}}, \citenamefont {{Farhi}}, \citenamefont {{Foxen}}, \citenamefont
		{{Giustina}}, \citenamefont {{Gross}}, \citenamefont {{Harrigan}},
		\citenamefont {{Harrington}}, \citenamefont {{Hilton}}, \citenamefont {{Ho}},
		\citenamefont {{Huang}}, \citenamefont {{Huggins}}, \citenamefont {{Ioffe}},
		\citenamefont {{Isakov}}, \citenamefont {{Jeffrey}}, \citenamefont {{Jiang}},
		\citenamefont {{Kechedzhi}}, \citenamefont {{Kim}}, \citenamefont
		{{Kostritsa}}, \citenamefont {{Landhuis}}, \citenamefont {{Laptev}},
		\citenamefont {{Lucero}}, \citenamefont {{Martin}}, \citenamefont
		{{McClean}}, \citenamefont {{McCourt}}, \citenamefont {{Mi}}, \citenamefont
		{{Miao}}, \citenamefont {{Mohseni}}, \citenamefont {{Mruczkiewicz}},
		\citenamefont {{Mutus}}, \citenamefont {{Naaman}}, \citenamefont {{Neeley}},
		\citenamefont {{Neill}}, \citenamefont {{Newman}}, \citenamefont {{Yuezhen
				Niu}}, \citenamefont {{O'Brien}}, \citenamefont {{Opremcak}}, \citenamefont
		{{Ostby}}, \citenamefont {{Pat{\'o}}}, \citenamefont {{Redd}}, \citenamefont
		{{Roushan}}, \citenamefont {{Rubin}}, \citenamefont {{Shvarts}},
		\citenamefont {{Strain}}, \citenamefont {{Szalay}}, \citenamefont
		{{Trevithick}}, \citenamefont {{Villalonga}}, \citenamefont {{White}},
		\citenamefont {{Yao}}, \citenamefont {{Yeh}}, \citenamefont {{Zalcman}},
		\citenamefont {{Neven}}, \citenamefont {{Boixo}}, \citenamefont
		{{Smelyanskiy}}, \citenamefont {{Chen}}, \citenamefont {{Megrant}},\ and\
		\citenamefont {{Kelly}}}]{Chen:21}%
	\BibitemOpen
	\bibfield  {author} {\bibinfo {author} {\bibfnamefont {Z.}~\bibnamefont
			{{Chen}}},  {\em et~al.},\ }\enquote{\bibinfo {title} {Exponential
			suppression of bit or phase errors with cyclic error correction}},\ \href
	{\doibase https://doi.org/10.1038/s41586-021-03588-y} {\bibfield  {journal}
		{\bibinfo  {journal} {Nature}\ }\textbf {\bibinfo {volume} {595}},\ \bibinfo
		{pages} {383} (\bibinfo {year} {2021})}\BibitemShut {NoStop}%
	\bibitem [{\citenamefont {Zhao}\ \textit
		{et~al.}(2022{\natexlab{b}})\citenamefont {Zhao}, \citenamefont {Ye},
		\citenamefont {Huang}, \citenamefont {Zhang}, \citenamefont {Wu},
		\citenamefont {Guan}, \citenamefont {Zhu}, \citenamefont {Wei}, \citenamefont
		{He}, \citenamefont {Cao}, \citenamefont {Chen}, \citenamefont {Chung},
		\citenamefont {Deng}, \citenamefont {Fan}, \citenamefont {Gong},
		\citenamefont {Guo}, \citenamefont {Guo}, \citenamefont {Han}, \citenamefont
		{Li}, \citenamefont {Li}, \citenamefont {Li}, \citenamefont {Liang},
		\citenamefont {Lin}, \citenamefont {Qian}, \citenamefont {Rong},
		\citenamefont {Su}, \citenamefont {Sun}, \citenamefont {Wang}, \citenamefont
		{Wu}, \citenamefont {Xu}, \citenamefont {Ying}, \citenamefont {Yu},
		\citenamefont {Zha}, \citenamefont {Zhang}, \citenamefont {Huo},
		\citenamefont {Lu}, \citenamefont {Peng}, \citenamefont {Zhu},\ and\
		\citenamefont {Pan}}]{Zhao:22}%
	\BibitemOpen
	\bibfield  {author} {\bibinfo {author} {\bibfnamefont {Y.}~\bibnamefont
			{Zhao}},  {\em et~al.},\ }\enquote{\bibinfo {title} {Realization of an
			Error-Correcting Surface Code with Superconducting Qubits}},\ \href {\doibase
		10.1103/PhysRevLett.129.030501} {\bibfield  {journal} {\bibinfo  {journal}
			{Phys. Rev. Lett.}\ }\textbf {\bibinfo {volume} {129}},\ \bibinfo {pages}
		{030501} (\bibinfo {year} {2022}{\natexlab{b}})}\BibitemShut {NoStop}%
	\bibitem [{\citenamefont {Mates}\ \textit {et~al.}(2019)\citenamefont {Mates},
		\citenamefont {Becker}, \citenamefont {Bennett}, \citenamefont {Dober},
		\citenamefont {Gard}, \citenamefont {Hilton}, \citenamefont {Swetz},
		\citenamefont {Vale},\ and\ \citenamefont {Ullom}}]{Mates:19}%
	\BibitemOpen
	\bibfield  {author} {\bibinfo {author} {\bibfnamefont {J.}~\bibnamefont
			{Mates}},  {\em et~al.},\ }\enquote{\bibinfo {title} {Crosstalk in microwave
			SQUID multiplexers}},\ \href {\doibase https://doi.org/10.1063/1.5116573}
	{\bibfield  {journal} {\bibinfo  {journal} {Applied Physics Letters}\
		}\textbf {\bibinfo {volume} {115}},\ \bibinfo {pages} {202601} (\bibinfo
		{year} {2019})}\BibitemShut {NoStop}%
	\bibitem [{\citenamefont {Mundada}\ \textit {et~al.}(2019)\citenamefont
		{Mundada}, \citenamefont {Zhang}, \citenamefont {Hazard},\ and\ \citenamefont
		{Houck}}]{Mundada:19}%
	\BibitemOpen
	\bibfield  {author} {\bibinfo {author} {\bibfnamefont {P.}~\bibnamefont
			{Mundada}}, \bibinfo {author} {\bibfnamefont {G.}~\bibnamefont {Zhang}},
		\bibinfo {author} {\bibfnamefont {T.}~\bibnamefont {Hazard}},  and \bibinfo
		{author} {\bibfnamefont {A.}~\bibnamefont {Houck}},\ }\enquote{\bibinfo
		{title} {Suppression of Qubit Crosstalk in a Tunable Coupling Superconducting
			Circuit}},\ \href {\doibase 10.1103/PhysRevApplied.12.054023} {\bibfield
		{journal} {\bibinfo  {journal} {Phys. Rev. Applied}\ }\textbf {\bibinfo
			{volume} {12}},\ \bibinfo {pages} {054023} (\bibinfo {year}
		{2019})}\BibitemShut {NoStop}%
	\bibitem [{\citenamefont {Dai}\ \textit {et~al.}(2021)\citenamefont {Dai},
		\citenamefont {Tennant}, \citenamefont {Trappen}, \citenamefont {Martinez},
		\citenamefont {Melanson}, \citenamefont {Yurtalan}, \citenamefont {Tang},
		\citenamefont {Novikov}, \citenamefont {Grover}, \citenamefont {Disseler},
		\citenamefont {Basham}, \citenamefont {Das}, \citenamefont {Kim},
		\citenamefont {Melville}, \citenamefont {Niedzielski}, \citenamefont {Weber},
		\citenamefont {Yoder}, \citenamefont {Lidar},\ and\ \citenamefont
		{Lupascu}}]{Dai:21}%
	\BibitemOpen
	\bibfield  {author} {\bibinfo {author} {\bibfnamefont {X.}~\bibnamefont
			{Dai}},  {\em et~al.},\ }\enquote{\bibinfo {title} {Calibration of Flux
			Crosstalk in Large-Scale Flux-Tunable Superconducting Quantum Circuits}},\
	\href {\doibase 10.1103/PRXQuantum.2.040313} {\bibfield  {journal} {\bibinfo
			{journal} {PRX Quantum}\ }\textbf {\bibinfo {volume} {2}},\ \bibinfo {pages}
		{040313} (\bibinfo {year} {2021})}\BibitemShut {NoStop}%
	\bibitem [{\citenamefont {Ni}\ \textit {et~al.}(2022)\citenamefont {Ni},
		\citenamefont {Li}, \citenamefont {Zhang}, \citenamefont {Chu}, \citenamefont
		{Niu}, \citenamefont {Yan}, \citenamefont {Deng}, \citenamefont {Hu},
		\citenamefont {Li}, \citenamefont {Zhong}, \citenamefont {Liu}, \citenamefont
		{Yan}, \citenamefont {Xu},\ and\ \citenamefont {Yu}}]{Ni:21}%
	\BibitemOpen
	\bibfield  {author} {\bibinfo {author} {\bibfnamefont {Z.}~\bibnamefont
			{Ni}},  {\em et~al.},\ }\enquote{\bibinfo {title} {Scalable Method for
			Eliminating Residual $ZZ$ Interaction between Superconducting Qubits}},\
	\href {\doibase 10.1103/PhysRevLett.129.040502} {\bibfield  {journal}
		{\bibinfo  {journal} {Phys. Rev. Lett.}\ }\textbf {\bibinfo {volume} {129}},\
		\bibinfo {pages} {040502} (\bibinfo {year} {2022})}\BibitemShut {NoStop}%
	\bibitem [{\citenamefont {Cai}\ \textit
		{et~al.}(2021{\natexlab{a}})\citenamefont {Cai}, \citenamefont {Han},
		\citenamefont {Wu}, \citenamefont {Ma}, \citenamefont {Wang}, \citenamefont
		{Wang}, \citenamefont {Zhang}, \citenamefont {Wang}, \citenamefont {Song},\
		and\ \citenamefont {Duan}}]{Cai:21}%
	\BibitemOpen
	\bibfield  {author} {\bibinfo {author} {\bibfnamefont {T.-Q.}\ \bibnamefont
			{Cai}},  {\em et~al.},\ }\enquote{\bibinfo {title} {Impact of Spectators on a
			Two-Qubit Gate in a Tunable Coupling Superconducting Circuit}},\ \href
	{\doibase 10.1103/PhysRevLett.127.060505} {\bibfield  {journal} {\bibinfo
			{journal} {Phys. Rev. Lett.}\ }\textbf {\bibinfo {volume} {127}},\ \bibinfo
		{pages} {060505} (\bibinfo {year} {2021}{\natexlab{a}})}\BibitemShut
	{NoStop}%
	\bibitem [{\citenamefont {Nuerbolati}\ \textit {et~al.}(2022)\citenamefont
		{Nuerbolati}, \citenamefont {Han}, \citenamefont {Chu}, \citenamefont {Zhou},
		\citenamefont {Tan}, \citenamefont {Yu}, \citenamefont {Liu},\ and\
		\citenamefont {Yan}}]{Nuerbolati:22}%
	\BibitemOpen
	\bibfield  {author} {\bibinfo {author} {\bibfnamefont {W.}~\bibnamefont
			{Nuerbolati}},  {\em et~al.},\ }\enquote{\bibinfo {title} {Canceling
			microwave crosstalk with fixed-frequency qubits}},\ \href {\doibase
		10.1063/5.0088094} {\bibfield  {journal} {\bibinfo  {journal} {Applied
				Physics Letters}\ }\textbf {\bibinfo {volume} {120}},\ \bibinfo {pages}
		{174001} (\bibinfo {year} {2022})}\BibitemShut {NoStop}%
	\bibitem [{\citenamefont {Sheldon}\ \textit {et~al.}(2016)\citenamefont
		{Sheldon}, \citenamefont {Magesan}, \citenamefont {Chow},\ and\ \citenamefont
		{Gambetta}}]{Sheldon:16}%
	\BibitemOpen
	\bibfield  {author} {\bibinfo {author} {\bibfnamefont {S.}~\bibnamefont
			{Sheldon}}, \bibinfo {author} {\bibfnamefont {E.}~\bibnamefont {Magesan}},
		\bibinfo {author} {\bibfnamefont {J.~M.}\ \bibnamefont {Chow}},  and \bibinfo
		{author} {\bibfnamefont {J.~M.}\ \bibnamefont {Gambetta}},\
	}\enquote{\bibinfo {title} {Procedure for systematically tuning up cross-talk
			in the cross-resonance gate}},\ \href {\doibase 10.1103/PhysRevA.93.060302}
	{\bibfield  {journal} {\bibinfo  {journal} {Phys. Rev. A}\ }\textbf {\bibinfo
			{volume} {93}},\ \bibinfo {pages} {060302} (\bibinfo {year}
		{2016})}\BibitemShut {NoStop}%
	\bibitem [{\citenamefont {Blais}\ \textit {et~al.}(2004)\citenamefont {Blais},
		\citenamefont {Huang}, \citenamefont {Wallraff}, \citenamefont {Girvin},\
		and\ \citenamefont {Schoelkopf}}]{Blais:04}%
	\BibitemOpen
	\bibfield  {author} {\bibinfo {author} {\bibfnamefont {A.}~\bibnamefont
			{Blais}},  {\em et~al.},\ }\enquote{\bibinfo {title} {Cavity quantum
			electrodynamics for superconducting electrical circuits: An architecture for
			quantum computation}},\ \href {\doibase 10.1103/PhysRevA.69.062320}
	{\bibfield  {journal} {\bibinfo  {journal} {Phys. Rev. A}\ }\textbf {\bibinfo
			{volume} {69}},\ \bibinfo {pages} {062320} (\bibinfo {year}
		{2004})}\BibitemShut {NoStop}%
	\bibitem [{\citenamefont {Blais}\ \textit {et~al.}(2007)\citenamefont {Blais},
		\citenamefont {Gambetta}, \citenamefont {Wallraff}, \citenamefont {Schuster},
		\citenamefont {Girvin}, \citenamefont {Devoret},\ and\ \citenamefont
		{Schoelkopf}}]{Blais:07}%
	\BibitemOpen
	\bibfield  {author} {\bibinfo {author} {\bibfnamefont {A.}~\bibnamefont
			{Blais}},  {\em et~al.},\ }\enquote{\bibinfo {title} {Quantum-information
			processing with circuit quantum electrodynamics}},\ \href {\doibase
		10.1103/PhysRevA.75.032329} {\bibfield  {journal} {\bibinfo  {journal} {Phys.
				Rev. A}\ }\textbf {\bibinfo {volume} {75}},\ \bibinfo {pages} {032329}
		(\bibinfo {year} {2007})}\BibitemShut {NoStop}%
	\bibitem [{\citenamefont {Wallraff}\ \textit {et~al.}(2004)\citenamefont
		{Wallraff}, \citenamefont {Schuster}, \citenamefont {Blais}, \citenamefont
		{Frunzio}, \citenamefont {Huang}, \citenamefont {Majer}, \citenamefont
		{Kumar}, \citenamefont {Girvin},\ and\ \citenamefont
		{Schoelkopf}}]{Wallraff:04}%
	\BibitemOpen
	\bibfield  {author} {\bibinfo {author} {\bibfnamefont {A.}~\bibnamefont
			{Wallraff}},  {\em et~al.},\ }\enquote{\bibinfo {title} {Strong coupling of a
			single photon to a superconducting qubit using circuit quantum
			electrodynamics}},\ \href {https://www.nature.com/articles/nature02851}
	{\bibfield  {journal} {\bibinfo  {journal} {Nature}\ }\textbf {\bibinfo
			{volume} {431}},\ \bibinfo {pages} {162} (\bibinfo {year}
		{2004})}\BibitemShut {NoStop}%
	\bibitem [{\citenamefont {{Majer}}\ \textit {et~al.}(2007)\citenamefont
		{{Majer}}, \citenamefont {{Chow}}, \citenamefont {{Gambetta}}, \citenamefont
		{{Koch}}, \citenamefont {{Johnson}}, \citenamefont {{Schreier}},
		\citenamefont {{Frunzio}}, \citenamefont {{Schuster}}, \citenamefont
		{{Houck}}, \citenamefont {{Wallraff}}, \citenamefont {{Blais}}, \citenamefont
		{{Devoret}}, \citenamefont {{Girvin}},\ and\ \citenamefont
		{{Schoelkopf}}}]{Majer:07}%
	\BibitemOpen
	\bibfield  {author} {\bibinfo {author} {\bibfnamefont {J.}~\bibnamefont
			{{Majer}}},  {\em et~al.},\ }\enquote{\bibinfo {title} {Coupling
			superconducting qubits via a cavity bus}},\ \href {\doibase
		https://doi.org/10.1038/nature06184} {\bibfield  {journal} {\bibinfo
			{journal} {Nature}\ }\textbf {\bibinfo {volume} {449}},\ \bibinfo {pages}
		{443} (\bibinfo {year} {2007})}\BibitemShut {NoStop}%
	\bibitem [{\citenamefont {Chow}\ \textit {et~al.}(2011)\citenamefont {Chow},
		\citenamefont {C\'orcoles}, \citenamefont {Gambetta}, \citenamefont
		{Rigetti}, \citenamefont {Johnson}, \citenamefont {Smolin}, \citenamefont
		{Rozen}, \citenamefont {Keefe}, \citenamefont {Rothwell}, \citenamefont
		{Ketchen},\ and\ \citenamefont {Steffen}}]{Chow:11}%
	\BibitemOpen
	\bibfield  {author} {\bibinfo {author} {\bibfnamefont {J.~M.}\ \bibnamefont
			{Chow}},  {\em et~al.},\ }\enquote{\bibinfo {title} {Simple All-Microwave
			Entangling Gate for Fixed-Frequency Superconducting Qubits}},\ \href
	{\doibase 10.1103/PhysRevLett.107.080502} {\bibfield  {journal} {\bibinfo
			{journal} {Phys. Rev. Lett.}\ }\textbf {\bibinfo {volume} {107}},\ \bibinfo
		{pages} {080502} (\bibinfo {year} {2011})}\BibitemShut {NoStop}%
	\bibitem [{\citenamefont {DiCarlo}\ \textit {et~al.}(2009)\citenamefont
		{DiCarlo}, \citenamefont {Chow}, \citenamefont {Gambetta}, \citenamefont
		{Bishop}, \citenamefont {Johnson}, \citenamefont {Schuster}, \citenamefont
		{Majer}, \citenamefont {Blais}, \citenamefont {Frunzio}, \citenamefont
		{Girvin} \textit {et~al.}}]{DiCarlo:09}%
	\BibitemOpen
	\bibfield  {author} {\bibinfo {author} {\bibfnamefont {L.}~\bibnamefont
			{DiCarlo}},  {\em et~al.},\ }\enquote{\bibinfo {title} {Demonstration of
			two-qubit algorithms with a superconducting quantum processor}},\ \href
	{\doibase https://doi.org/10.1038/nature08121} {\bibfield  {journal}
		{\bibinfo  {journal} {Nature}\ }\textbf {\bibinfo {volume} {460}},\ \bibinfo
		{pages} {240} (\bibinfo {year} {2009})}\BibitemShut {NoStop}%
	\bibitem [{\citenamefont {Johnson}\ \textit {et~al.}(2010)\citenamefont
		{Johnson}, \citenamefont {Reed}, \citenamefont {Houck}, \citenamefont
		{Schuster}, \citenamefont {Bishop}, \citenamefont {Ginossar}, \citenamefont
		{Gambetta}, \citenamefont {DiCarlo}, \citenamefont {Frunzio}, \citenamefont
		{Girvin} \textit {et~al.}}]{Johnson:10}%
	\BibitemOpen
	\bibfield  {author} {\bibinfo {author} {\bibfnamefont {B.}~\bibnamefont
			{Johnson}},  {\em et~al.},\ }\enquote{\bibinfo {title} {Quantum
			non-demolition detection of single microwave photons in a circuit}},\ \href
	{\doibase https://doi.org/10.1038/nphys1710} {\bibfield  {journal} {\bibinfo
			{journal} {Nature Physics}\ }\textbf {\bibinfo {volume} {6}},\ \bibinfo
		{pages} {663} (\bibinfo {year} {2010})}\BibitemShut {NoStop}%
	\bibitem [{\citenamefont {Leek}\ \textit {et~al.}(2010)\citenamefont {Leek},
		\citenamefont {Baur}, \citenamefont {Fink}, \citenamefont {Bianchetti},
		\citenamefont {Steffen}, \citenamefont {Filipp},\ and\ \citenamefont
		{Wallraff}}]{Leek:10}%
	\BibitemOpen
	\bibfield  {author} {\bibinfo {author} {\bibfnamefont {P.~J.}\ \bibnamefont
			{Leek}},  {\em et~al.},\ }\enquote{\bibinfo {title} {Cavity Quantum
			Electrodynamics with Separate Photon Storage and Qubit Readout Modes}},\
	\href {\doibase 10.1103/PhysRevLett.104.100504} {\bibfield  {journal}
		{\bibinfo  {journal} {Phys. Rev. Lett.}\ }\textbf {\bibinfo {volume} {104}},\
		\bibinfo {pages} {100504} (\bibinfo {year} {2010})}\BibitemShut {NoStop}%
	\bibitem [{\citenamefont {McKay}\ \textit {et~al.}(2019)\citenamefont {McKay},
		\citenamefont {Sheldon}, \citenamefont {Smolin}, \citenamefont {Chow},\ and\
		\citenamefont {Gambetta}}]{McKay:19}%
	\BibitemOpen
	\bibfield  {author} {\bibinfo {author} {\bibfnamefont {D.~C.}\ \bibnamefont
			{McKay}},  {\em et~al.},\ }\enquote{\bibinfo {title} {Three-Qubit Randomized
			Benchmarking}},\ \href {\doibase 10.1103/PhysRevLett.122.200502} {\bibfield
		{journal} {\bibinfo  {journal} {Phys. Rev. Lett.}\ }\textbf {\bibinfo
			{volume} {122}},\ \bibinfo {pages} {200502} (\bibinfo {year}
		{2019})}\BibitemShut {NoStop}%
	\bibitem [{\citenamefont {Cai}\ \textit
		{et~al.}(2021{\natexlab{b}})\citenamefont {Cai}, \citenamefont {Wang},
		\citenamefont {Wang}, \citenamefont {Han}, \citenamefont {Wu}, \citenamefont
		{Song},\ and\ \citenamefont {Duan}}]{Cai:21PRR}%
	\BibitemOpen
	\bibfield  {author} {\bibinfo {author} {\bibfnamefont {T.-Q.}\ \bibnamefont
			{Cai}},  {\em et~al.},\ }\enquote{\bibinfo {title} {All-microwave
			nonadiabatic multiqubit geometric phase gate for superconducting qubits}},\
	\href {\doibase 10.1103/PhysRevResearch.3.043071} {\bibfield  {journal}
		{\bibinfo  {journal} {Phys. Rev. Research}\ }\textbf {\bibinfo {volume}
			{3}},\ \bibinfo {pages} {043071} (\bibinfo {year}
		{2021}{\natexlab{b}})}\BibitemShut {NoStop}%
	\bibitem [{\citenamefont {Fink}\ \textit {et~al.}(2009)\citenamefont {Fink},
		\citenamefont {Bianchetti}, \citenamefont {Baur}, \citenamefont {G\"oppl},
		\citenamefont {Steffen}, \citenamefont {Filipp}, \citenamefont {Leek},
		\citenamefont {Blais},\ and\ \citenamefont {Wallraff}}]{Fink:09}%
	\BibitemOpen
	\bibfield  {author} {\bibinfo {author} {\bibfnamefont {J.~M.}\ \bibnamefont
			{Fink}},  {\em et~al.},\ }\enquote{\bibinfo {title} {Dressed Collective Qubit
			States and the Tavis-Cummings Model in Circuit QED}},\ \href {\doibase
		10.1103/PhysRevLett.103.083601} {\bibfield  {journal} {\bibinfo  {journal}
			{Phys. Rev. Lett.}\ }\textbf {\bibinfo {volume} {103}},\ \bibinfo {pages}
		{083601} (\bibinfo {year} {2009})}\BibitemShut {NoStop}%
	\bibitem [{\citenamefont {Wang}\ \textit {et~al.}(2020)\citenamefont {Wang},
		\citenamefont {Li}, \citenamefont {Feng}, \citenamefont {Song}, \citenamefont
		{Song}, \citenamefont {Liu}, \citenamefont {Guo}, \citenamefont {Zhang},
		\citenamefont {Dong}, \citenamefont {Zheng}, \citenamefont {Wang},\ and\
		\citenamefont {Wang}}]{Wang:20}%
	\BibitemOpen
	\bibfield  {author} {\bibinfo {author} {\bibfnamefont {Z.}~\bibnamefont
			{Wang}},  {\em et~al.},\ }\enquote{\bibinfo {title} {Controllable Switching
			between Superradiant and Subradiant States in a 10-qubit Superconducting
			Circuit}},\ \href {\doibase 10.1103/PhysRevLett.124.013601} {\bibfield
		{journal} {\bibinfo  {journal} {Phys. Rev. Lett.}\ }\textbf {\bibinfo
			{volume} {124}},\ \bibinfo {pages} {013601} (\bibinfo {year}
		{2020})}\BibitemShut {NoStop}%
	\bibitem [{\citenamefont {Blais}\ \textit {et~al.}(2021)\citenamefont {Blais},
		\citenamefont {Grimsmo}, \citenamefont {Girvin},\ and\ \citenamefont
		{Wallraff}}]{Blais21}%
	\BibitemOpen
	\bibfield  {author} {\bibinfo {author} {\bibfnamefont {A.}~\bibnamefont
			{Blais}}, \bibinfo {author} {\bibfnamefont {A.~L.}\ \bibnamefont {Grimsmo}},
		\bibinfo {author} {\bibfnamefont {S.~M.}\ \bibnamefont {Girvin}},  and
		\bibinfo {author} {\bibfnamefont {A.}~\bibnamefont {Wallraff}},\
	}\enquote{\bibinfo {title} {Circuit quantum electrodynamics}},\ \href
	{\doibase 10.1103/RevModPhys.93.025005} {\bibfield  {journal} {\bibinfo
			{journal} {Rev. Mod. Phys.}\ }\textbf {\bibinfo {volume} {93}},\ \bibinfo
		{pages} {025005} (\bibinfo {year} {2021})}\BibitemShut {NoStop}%
	\bibitem [{\citenamefont {Berke}\ \textit {et~al.}(2022)\citenamefont {Berke},
		\citenamefont {Varvelis}, \citenamefont {Trebst}, \citenamefont {Altland},\
		and\ \citenamefont {DiVincenzo}}]{Berke:22}%
	\BibitemOpen
	\bibfield  {author} {\bibinfo {author} {\bibfnamefont {C.}~\bibnamefont
			{Berke}},  {\em et~al.},\ }\enquote{\bibinfo {title} {Transmon platform for
			quantum computing challenged by chaotic fluctuations}},\ \href {\doibase
		https://doi.org/10.1038/s41467-022-29940-y} {\bibfield  {journal} {\bibinfo
			{journal} {Nature communications}\ }\textbf {\bibinfo {volume} {13}},\
		\bibinfo {pages} {1} (\bibinfo {year} {2022})}\BibitemShut {NoStop}%
	\bibitem [{\citenamefont {Xu}\ \textit {et~al.}(2020)\citenamefont {Xu},
		\citenamefont {Chu}, \citenamefont {Yuan}, \citenamefont {Qiu}, \citenamefont
		{Zhou}, \citenamefont {Zhang}, \citenamefont {Tan}, \citenamefont {Yu},
		\citenamefont {Liu}, \citenamefont {Li}, \citenamefont {Yan},\ and\
		\citenamefont {Yu}}]{Xu:20}%
	\BibitemOpen
	\bibfield  {author} {\bibinfo {author} {\bibfnamefont {Y.}~\bibnamefont
			{Xu}},  {\em et~al.},\ }\enquote{\bibinfo {title} {High-Fidelity,
			High-Scalability Two-Qubit Gate Scheme for Superconducting Qubits}},\ \href
	{\doibase 10.1103/PhysRevLett.125.240503} {\bibfield  {journal} {\bibinfo
			{journal} {Phys. Rev. Lett.}\ }\textbf {\bibinfo {volume} {125}},\ \bibinfo
		{pages} {240503} (\bibinfo {year} {2020})}\BibitemShut {NoStop}%
	\bibitem [{\citenamefont {{Chu}}\ \textit {et~al.}(2022)\citenamefont {{Chu}},
		\citenamefont {{He}}, \citenamefont {{Zhou}}, \citenamefont {{Yuan}},
		\citenamefont {{Zhang}}, \citenamefont {{Guo}}, \citenamefont {{Hai}},
		\citenamefont {{Han}}, \citenamefont {{Hu}}, \citenamefont {{Huang}},
		\citenamefont {{Jia}}, \citenamefont {{Jiao}}, \citenamefont {{Liu}},
		\citenamefont {{Ni}}, \citenamefont {{Pan}}, \citenamefont {{Qiu}},
		\citenamefont {{Wei}}, \citenamefont {{Yang}}, \citenamefont {{Zhang}},
		\citenamefont {{Zhang}}, \citenamefont {{Zou}}, \citenamefont {{Chen}},
		\citenamefont {{Deng}}, \citenamefont {{Deng}}, \citenamefont {{Hu}},
		\citenamefont {{Li}}, \citenamefont {{Tan}}, \citenamefont {{Xu}},
		\citenamefont {{Yan}}, \citenamefont {{Sun}}, \citenamefont {{Yan}},\ and\
		\citenamefont {{Yu}}}]{Chu:22}%
	\BibitemOpen
	\bibfield  {author} {\bibinfo {author} {\bibfnamefont {J.}~\bibnamefont
			{{Chu}}},  {\em et~al.},\ }\enquote{\bibinfo {title} {{Scalable algorithm
				simplification using quantum AND logic}}},\ \href {\doibase
		10.1038/s41567-022-01813-7} {\bibfield  {journal} {\bibinfo  {journal}
			{Nature Physics}\ } (\bibinfo {year} {2022}),\
		10.1038/s41567-022-01813-7}\BibitemShut {NoStop}%
	\bibitem [{\citenamefont {{Luo}}\ \textit {et~al.}(2022)\citenamefont {{Luo}},
		\citenamefont {{Huang}}, \citenamefont {{Tao}}, \citenamefont {{Zhang}},
		\citenamefont {{Zhou}}, \citenamefont {{Chu}}, \citenamefont {{Liu}},
		\citenamefont {{Wang}}, \citenamefont {{Cui}}, \citenamefont {{Liu}},
		\citenamefont {{Yan}}, \citenamefont {{Yung}}, \citenamefont {{Chen}},
		\citenamefont {{Yan}},\ and\ \citenamefont {{Yu}}}]{Luo:22}%
	\BibitemOpen
	\bibfield  {author} {\bibinfo {author} {\bibfnamefont {K.}~\bibnamefont
			{{Luo}}},  {\em et~al.},\ }\enquote{\bibinfo {title} {{Experimental
				Realization of Two Qutrits Gate with Tunable Coupling in Superconducting
				Circuits}}},\ \href@noop {} {\bibfield  {journal} {\bibinfo  {journal} {arXiv
				e-prints}\ ,\ \bibinfo {eid} {arXiv:2206.11199}} (\bibinfo {year} {2022})},\
	\Eprint {http://arxiv.org/abs/2206.11199} {arXiv:2206.11199 [quant-ph]}
	\BibitemShut {NoStop}%
	\bibitem [{\citenamefont {{Hu}}\ \textit {et~al.}(2022)\citenamefont {{Hu}},
		\citenamefont {{Yuan}}, \citenamefont {{Veloso}}, \citenamefont {{Qiu}},
		\citenamefont {{Zhou}}, \citenamefont {{Zhang}}, \citenamefont {{Chu}},
		\citenamefont {{Nurbolat}}, \citenamefont {{Hu}}, \citenamefont {{Li}},
		\citenamefont {{Xu}}, \citenamefont {{Zhong}}, \citenamefont {{Liu}},
		\citenamefont {{Yan}}, \citenamefont {{Tan}}, \citenamefont {{Bachelard}},
		\citenamefont {{Santos}}, \citenamefont {{Villas-Boas}},\ and\ \citenamefont
		{{Yu}}}]{Bruno:22}%
	\BibitemOpen
	\bibfield  {author} {\bibinfo {author} {\bibfnamefont {C.-K.}\ \bibnamefont
			{{Hu}}},  {\em et~al.},\ }\enquote{\bibinfo {title} {{Conditional coherent
				control with superconducting artificial atoms}}},\ \href@noop {} {\bibfield
		{journal} {\bibinfo  {journal} {arXiv e-prints}\ ,\ \bibinfo {eid}
			{arXiv:2203.09791}} (\bibinfo {year} {2022})},\ \Eprint
	{http://arxiv.org/abs/2203.09791} {arXiv:2203.09791 [quant-ph]} \BibitemShut
	{NoStop}%
	\bibitem [{\citenamefont {Mitchell}\ \textit {et~al.}(2021)\citenamefont
		{Mitchell}, \citenamefont {Naik}, \citenamefont {Morvan}, \citenamefont
		{Hashim}, \citenamefont {Kreikebaum}, \citenamefont {Marinelli},
		\citenamefont {Lavrijsen}, \citenamefont {Nowrouzi}, \citenamefont
		{Santiago},\ and\ \citenamefont {Siddiqi}}]{Mitchell:21}%
	\BibitemOpen
	\bibfield  {author} {\bibinfo {author} {\bibfnamefont {B.~K.}\ \bibnamefont
			{Mitchell}},  {\em et~al.},\ }\enquote{\bibinfo {title} {Hardware-Efficient
			Microwave-Activated Tunable Coupling between Superconducting Qubits}},\ \href
	{\doibase 10.1103/PhysRevLett.127.200502} {\bibfield  {journal} {\bibinfo
			{journal} {Phys. Rev. Lett.}\ }\textbf {\bibinfo {volume} {127}},\ \bibinfo
		{pages} {200502} (\bibinfo {year} {2021})}\BibitemShut {NoStop}%
	\bibitem [{\citenamefont {Klink}(1997)}]{Klink:97}%
	\BibitemOpen
	\bibfield  {author} {\bibinfo {author} {\bibfnamefont {W.}~\bibnamefont
			{Klink}},\ }\enquote{\bibinfo {title} {Quantum Mechanics in Nonintertial
			Reference Frames}},\ \href {\doibase https://doi.org/10.1006/aphy.1997.5720}
	{\bibfield  {journal} {\bibinfo  {journal} {Annals of Physics}\ }\textbf
		{\bibinfo {volume} {260}},\ \bibinfo {pages} {27 } (\bibinfo {year}
		{1997})}\BibitemShut {NoStop}%
	\bibitem [{\citenamefont {Serov}(2017)}]{Serov:17}%
	\BibitemOpen
	\bibfield  {author} {\bibinfo {author} {\bibfnamefont {V.}~\bibnamefont
			{Serov}},\ }\enquote{\bibinfo {title} {The Riemann--Lebesgue Lemma}},\ in\
	\href {\doibase 10.1007/978-3-319-65262-7_6} {\textit {\bibinfo {booktitle}
			{Fourier Series, Fourier Transform and Their Applications to Mathematical
				Physics}}}\ (\bibinfo  {publisher} {Springer International Publishing},\
	\bibinfo {address} {Cham},\ \bibinfo {year} {2017})\ pp.\ \bibinfo {pages}
	{33--35}\BibitemShut {NoStop}%
	\bibitem [{\citenamefont {Bravyi}\ \textit {et~al.}(2011)\citenamefont
		{Bravyi}, \citenamefont {DiVincenzo},\ and\ \citenamefont
		{Loss}}]{Bravyi:11}%
	\BibitemOpen
	\bibfield  {author} {\bibinfo {author} {\bibfnamefont {S.}~\bibnamefont
			{Bravyi}}, \bibinfo {author} {\bibfnamefont {D.~P.}\ \bibnamefont
			{DiVincenzo}},  and \bibinfo {author} {\bibfnamefont {D.}~\bibnamefont
			{Loss}},\ }\enquote{\bibinfo {title} {Schrieffer–Wolff transformation for
			quantum many-body systems}},\ \href {\doibase
		https://doi.org/10.1016/j.aop.2011.06.004} {\bibfield  {journal} {\bibinfo
			{journal} {Annals of Physics}\ }\textbf {\bibinfo {volume} {326}},\ \bibinfo
		{pages} {2793 } (\bibinfo {year} {2011})}\BibitemShut {NoStop}%
	\bibitem [{\citenamefont {Yan}\ \textit {et~al.}(2018)\citenamefont {Yan},
		\citenamefont {Krantz}, \citenamefont {Sung}, \citenamefont {Kjaergaard},
		\citenamefont {Campbell}, \citenamefont {Orlando}, \citenamefont
		{Gustavsson},\ and\ \citenamefont {Oliver}}]{Yan:18}%
	\BibitemOpen
	\bibfield  {author} {\bibinfo {author} {\bibfnamefont {F.}~\bibnamefont
			{Yan}},  {\em et~al.},\ }\enquote{\bibinfo {title} {Tunable Coupling Scheme
			for Implementing High-Fidelity Two-Qubit Gates}},\ \href {\doibase
		10.1103/PhysRevApplied.10.054062} {\bibfield  {journal} {\bibinfo  {journal}
			{Phys. Rev. Applied}\ }\textbf {\bibinfo {volume} {10}},\ \bibinfo {pages}
		{054062} (\bibinfo {year} {2018})}\BibitemShut {NoStop}%
	\bibitem [{\citenamefont {Li}\ \textit {et~al.}(2020)\citenamefont {Li},
		\citenamefont {Cai}, \citenamefont {Yan}, \citenamefont {Wang}, \citenamefont
		{Pan}, \citenamefont {Ma}, \citenamefont {Cai}, \citenamefont {Han},
		\citenamefont {Hua}, \citenamefont {Han}, \citenamefont {Wu}, \citenamefont
		{Zhang}, \citenamefont {Wang}, \citenamefont {Song}, \citenamefont {Duan},\
		and\ \citenamefont {Sun}}]{Li:20}%
	\BibitemOpen
	\bibfield  {author} {\bibinfo {author} {\bibfnamefont {X.}~\bibnamefont
			{Li}},  {\em et~al.},\ }\enquote{\bibinfo {title} {Tunable Coupler for
			Realizing a Controlled-Phase Gate with Dynamically Decoupled Regime in a
			Superconducting Circuit}},\ \href {\doibase 10.1103/PhysRevApplied.14.024070}
	{\bibfield  {journal} {\bibinfo  {journal} {Phys. Rev. Applied}\ }\textbf
		{\bibinfo {volume} {14}},\ \bibinfo {pages} {024070} (\bibinfo {year}
		{2020})}\BibitemShut {NoStop}%
	\bibitem [{\citenamefont {Feng}\ and\ \citenamefont {Wang}(2020)}]{Feng:20}%
	\BibitemOpen
	\bibfield  {author} {\bibinfo {author} {\bibfnamefont {W.}~\bibnamefont
			{Feng}} and \bibinfo {author} {\bibfnamefont {D.-w.}\ \bibnamefont {Wang}},\
	}\enquote{\bibinfo {title} {Quantum Fredkin gate based on synthetic
			three-body interactions in superconducting circuits}},\ \href {\doibase
		10.1103/PhysRevA.101.062312} {\bibfield  {journal} {\bibinfo  {journal}
			{Phys. Rev. A}\ }\textbf {\bibinfo {volume} {101}},\ \bibinfo {pages}
		{062312} (\bibinfo {year} {2020})}\BibitemShut {NoStop}%
	\bibitem [{\citenamefont {Johansson}\ \textit {et~al.}(2012)\citenamefont
		{Johansson}, \citenamefont {Nation},\ and\ \citenamefont {Nori}}]{QuTiP-1}%
	\BibitemOpen
	\bibfield  {author} {\bibinfo {author} {\bibfnamefont {J.}~\bibnamefont
			{Johansson}}, \bibinfo {author} {\bibfnamefont {P.}~\bibnamefont {Nation}},
		and \bibinfo {author} {\bibfnamefont {F.}~\bibnamefont {Nori}},\
	}\enquote{\bibinfo {title} {QuTiP: An open-source Python framework for the
			dynamics of open quantum systems}},\ \href {\doibase
		https://doi.org/10.1016/j.cpc.2012.02.021} {\bibfield  {journal} {\bibinfo
			{journal} {Computer Physics Communications}\ }\textbf {\bibinfo {volume}
			{183}},\ \bibinfo {pages} {1760} (\bibinfo {year} {2012})}\BibitemShut
	{NoStop}%
	\bibitem [{\citenamefont {Johansson}\ \textit {et~al.}(2013)\citenamefont
		{Johansson}, \citenamefont {Nation},\ and\ \citenamefont {Nori}}]{QuTiP-2}%
	\BibitemOpen
	\bibfield  {author} {\bibinfo {author} {\bibfnamefont {J.}~\bibnamefont
			{Johansson}}, \bibinfo {author} {\bibfnamefont {P.}~\bibnamefont {Nation}},
		and \bibinfo {author} {\bibfnamefont {F.}~\bibnamefont {Nori}},\
	}\enquote{\bibinfo {title} {QuTiP 2: A Python framework for the dynamics of
			open quantum systems}},\ \href {\doibase
		https://doi.org/10.1016/j.cpc.2012.11.019} {\bibfield  {journal} {\bibinfo
			{journal} {Computer Physics Communications}\ }\textbf {\bibinfo {volume}
			{184}},\ \bibinfo {pages} {1234} (\bibinfo {year} {2013})}\BibitemShut
	{NoStop}%
	\bibitem [{\citenamefont {Nielsen}\ and\ \citenamefont
		{Chuang}(2011)}]{Nielsen:Book}%
	\BibitemOpen
	\bibfield  {author} {\bibinfo {author} {\bibfnamefont {M.~A.}\ \bibnamefont
			{Nielsen}} and \bibinfo {author} {\bibfnamefont {I.~L.}\ \bibnamefont
			{Chuang}},\ }\href {\doibase 10.1017/CBO9780511976667} {\textit {\bibinfo
			{title} {Quantum Computation and Quantum Information: 10th Anniversary
				Edition}}},\ \bibinfo {edition} {10th}\ ed.\ (\bibinfo  {publisher}
	{Cambridge University Press},\ \bibinfo {address} {New York, NY, USA},\
	\bibinfo {year} {2011})\BibitemShut {NoStop}%
	\bibitem [{\citenamefont {Yurtalan}\ \textit {et~al.}(2020)\citenamefont
		{Yurtalan}, \citenamefont {Shi}, \citenamefont {Kononenko}, \citenamefont
		{Lupascu},\ and\ \citenamefont {Ashhab}}]{Yurtalan:20}%
	\BibitemOpen
	\bibfield  {author} {\bibinfo {author} {\bibfnamefont {M.~A.}\ \bibnamefont
			{Yurtalan}},  {\em et~al.},\ }\enquote{\bibinfo {title} {Implementation of a
			Walsh-Hadamard Gate in a Superconducting Qutrit}},\ \href {\doibase
		10.1103/PhysRevLett.125.180504} {\bibfield  {journal} {\bibinfo  {journal}
			{Phys. Rev. Lett.}\ }\textbf {\bibinfo {volume} {125}},\ \bibinfo {pages}
		{180504} (\bibinfo {year} {2020})}\BibitemShut {NoStop}%
	\bibitem [{\citenamefont {Kononenko}\ \textit {et~al.}(2021)\citenamefont
		{Kononenko}, \citenamefont {Yurtalan}, \citenamefont {Ren}, \citenamefont
		{Shi}, \citenamefont {Ashhab},\ and\ \citenamefont {Lupascu}}]{Kononenko:21}%
	\BibitemOpen
	\bibfield  {author} {\bibinfo {author} {\bibfnamefont {M.}~\bibnamefont
			{Kononenko}},  {\em et~al.},\ }\enquote{\bibinfo {title} {Characterization of
			control in a superconducting qutrit using randomized benchmarking}},\ \href
	{\doibase 10.1103/PhysRevResearch.3.L042007} {\bibfield  {journal} {\bibinfo
			{journal} {Phys. Rev. Research}\ }\textbf {\bibinfo {volume} {3}},\ \bibinfo
		{pages} {L042007} (\bibinfo {year} {2021})}\BibitemShut {NoStop}%
	\bibitem [{\citenamefont {{Goss}}\ \textit {et~al.}(2022)\citenamefont
		{{Goss}}, \citenamefont {{Morvan}}, \citenamefont {{Marinelli}},
		\citenamefont {{Mitchell}}, \citenamefont {{Nguyen}}, \citenamefont {{Naik}},
		\citenamefont {{Chen}}, \citenamefont {{J{\"u}nger}}, \citenamefont
		{{Kreikebaum}}, \citenamefont {{Santiago}}, \citenamefont {{Wallman}},\ and\
		\citenamefont {{Siddiqi}}}]{Goss:22}%
	\BibitemOpen
	\bibfield  {author} {\bibinfo {author} {\bibfnamefont {N.}~\bibnamefont
			{{Goss}}},  {\em et~al.},\ }\enquote{\bibinfo {title} {{High-Fidelity Qutrit
				Entangling Gates for Superconducting Circuits}}},\ \href {\doibase
		https://doi.org/10.1038/s41467-022-34851-z} {\bibfield  {journal} {\bibinfo
			{journal} {Nature Communications}\ }\textbf {\bibinfo {volume} {13}},\
		\bibinfo {pages} {1} (\bibinfo {year} {2022})}\BibitemShut {NoStop}%
	\bibitem [{\citenamefont {Xue}\ \textit {et~al.}(2017)\citenamefont {Xue},
		\citenamefont {Gu}, \citenamefont {Hong}, \citenamefont {Yang}, \citenamefont
		{Zhang}, \citenamefont {Hu},\ and\ \citenamefont {You}}]{Xue:17}%
	\BibitemOpen
	\bibfield  {author} {\bibinfo {author} {\bibfnamefont {Z.-Y.}\ \bibnamefont
			{Xue}},  {\em et~al.},\ }\enquote{\bibinfo {title} {Nonadiabatic Holonomic
			Quantum Computation with Dressed-State Qubits}},\ \href {\doibase
		10.1103/PhysRevApplied.7.054022} {\bibfield  {journal} {\bibinfo  {journal}
			{Phys. Rev. Applied}\ }\textbf {\bibinfo {volume} {7}},\ \bibinfo {pages}
		{054022} (\bibinfo {year} {2017})}\BibitemShut {NoStop}%
	\bibitem [{\citenamefont {Li}\ \textit {et~al.}(2021)\citenamefont {Li},
		\citenamefont {Liu}, \citenamefont {Ni}, \citenamefont {Zhang}, \citenamefont
		{Xue}, \citenamefont {Li}, \citenamefont {Yan}, \citenamefont {Chen},
		\citenamefont {Liu}, \citenamefont {Yung}, \citenamefont {Xu},\ and\
		\citenamefont {Yu}}]{Li:21}%
	\BibitemOpen
	\bibfield  {author} {\bibinfo {author} {\bibfnamefont {S.}~\bibnamefont
			{Li}},  {\em et~al.},\ }\enquote{\bibinfo {title} {Superrobust Geometric
			Control of a Superconducting Circuit}},\ \href {\doibase
		10.1103/PhysRevApplied.16.064003} {\bibfield  {journal} {\bibinfo  {journal}
			{Phys. Rev. Applied}\ }\textbf {\bibinfo {volume} {16}},\ \bibinfo {pages}
		{064003} (\bibinfo {year} {2021})}\BibitemShut {NoStop}%
	\bibitem [{\citenamefont {{Ma}}\ \textit {et~al.}(2022)\citenamefont {{Ma}},
		\citenamefont {{Xu}}, \citenamefont {{Chen}}, \citenamefont {{Zhang}},
		\citenamefont {{Zheng}}, \citenamefont {{Lan}}, \citenamefont {{Xue}},
		\citenamefont {{Tan}},\ and\ \citenamefont {{Yu}}}]{Ma:22}%
	\BibitemOpen
	\bibfield  {author} {\bibinfo {author} {\bibfnamefont {Z.}~\bibnamefont
			{{Ma}}},  {\em et~al.},\ }\enquote{\bibinfo {title} {{Experimental
				Implementation of Noncyclic and Nonadiabatic Geometric Quantum Gates in a
				Superconducting Circuit}}},\ \href@noop {} {\bibfield  {journal} {\bibinfo
			{journal} {arXiv e-prints}\ ,\ \bibinfo {eid} {arXiv:2210.03326}} (\bibinfo
		{year} {2022})},\ \Eprint {http://arxiv.org/abs/2210.03326} {arXiv:2210.03326
		[quant-ph]} \BibitemShut {NoStop}%
	\bibitem [{\citenamefont {Hu}\ \textit {et~al.}(2022)\citenamefont {Hu},
		\citenamefont {Qiu}, \citenamefont {Souza}, \citenamefont {Yuan},
		\citenamefont {Zhou}, \citenamefont {Zhang}, \citenamefont {Chu},
		\citenamefont {Pan}, \citenamefont {Hu}, \citenamefont {Li}, \citenamefont
		{Xu}, \citenamefont {Zhong}, \citenamefont {Liu}, \citenamefont {Yan},
		\citenamefont {Tan}, \citenamefont {Bachelard}, \citenamefont {Villas-Boas},
		\citenamefont {Santos},\ and\ \citenamefont {Yu}}]{Hu:22a}%
	\BibitemOpen
	\bibfield  {author} {\bibinfo {author} {\bibfnamefont {C.-K.}\ \bibnamefont
			{Hu}},  {\em et~al.},\ }\enquote{\bibinfo {title} {Optimal charging of a
			superconducting quantum battery}},\ \href {\doibase 10.1088/2058-9565/ac8444}
	{\bibfield  {journal} {\bibinfo  {journal} {Quantum Science and Technology}\
		}\textbf {\bibinfo {volume} {7}},\ \bibinfo {pages} {045018} (\bibinfo {year}
		{2022})}\BibitemShut {NoStop}%
\end{thebibliography}

%

\end{document}